\pdfoutput=1
\documentclass{aa}  
\usepackage{lscape}
\usepackage{graphicx}
\begin{document} 
   \title{Quasi-Stellar Objects acting as potential Strong Gravitational Lenses in the SDSS-III BOSS survey}
   \titlerunning{QSO Lenses in the BOSS survey}

   \author{Romain A. Meyer \inst{1,2} 
          \and Timoth\'ee Delubac \inst{1} \and Jean-Paul Kneib \inst{1}\fnmsep\inst{3} \and Fr\'ed\'eric Courbin\inst{1}
          }  
          
   \institute{Institute of Physics, Laboratory of Astrophysics,
Ecole Polytechnique F\'ed\'erale de Lausanne, CH-1015 Lausanne, Switzerland 
	\and
    Department of Physics and Astronomy, University College London, Gower Street, London WC1E 6BT, UK \\
       \email{r.meyer.17@ucl.ac.uk}
  	\and
    Aix Marseille Université, CNRS, LAM (Laboratoire d'Astrophysique de Marseille) UMR 7326, 13388, Marseille, France \\
         }
    \authorrunning{Meyer et al.}
             
   \date{Received 31/10/2017; Revised 08/03/2019 ; Accepted 11/03/2019}

\abstract{
We present a sample of $12$ Quasi-Stellar Objects (QSOs) potentially acting as strong gravitational lenses on background Emission Line Galaxies (ELG) or Lyman-$\alpha$ Emitters (LAEs) selected selected through a systematic search of the $297301$ QSOs in the Sloan Digital Sky Survey (SDSS)-III Data Release 12.
Candidates are identified by looking for compound spectra, where emission lines at a redshift larger than that of the quasar can be identified in the residuals after a QSO spectral template is subtracted from the observed spectra. The narrow diameter of BOSS fibers (2") then ensures that the object responsible for the additional emission lines must lie close to the line of sight of the QSO and hence provides a high probability of lensing.
Among the 12 candidates identified, $9$ have definite evidence for the presence of a background ELG identified by at least $4$ higher-redshift nebular emission lines. The remaining $3$ probable candidates present a strong asymmetrical emission line attributed to a background Lyman-$\alpha$ emitter (LAE). The QSO-ELG (QSO-LAE) lens candidates have QSO lens redshifts in the range $0.24\lesssim z_{\rm{QSO}} \lesssim 0.66$  ($0.75 \lesssim z_{\rm{QSO}} \lesssim 1.23$ ) and background galaxy redshifts in the range $0.48 \lesssim z_{S,ELG} \lesssim 0.94$ ($2.17 \lesssim z_{S,LAE} \lesssim 4.48$). We show that the algorithmic search is complete at $>90$ \% for QSO-ELG systems, whereas it falls at $40-60$\% for QSO-LAE, depending on the redshift of the source.
Upon confirmation of the lensing nature of the systems, this sample may quadruple the number of known QSOs acting as strong lenses. We have determined the completeness of our search, which allows future studies to compute lensing probabilities of galaxies by QSOs and differentiate between different QSO models. Future imaging of the full sample and lens modelling offers a unique approach to study and constrain key properties of QSOs.}

\keywords{gravitational lensing: strong --- methods: data analysis --- surveys: BOSS --- quasars: general}

\maketitle

\section{Introduction}
As gravitational lensing produces unmistakably distorted, amplified and multiplied images of the lensed objects, it is not surprising that the first gravitational lenses were discovered by identifying multiply imaged bright sources such as Quasi Stellar Objects (QSOs)~\citep[e.g.][]{Walsh1979,Weymann80,Young1981, Huchra1985}.
Since the first discoveries, the increasing number of wide-field surveys has revealed the use, first suggested by~\citet{Zwicky1937}, of strong gravitational lensing as a powerful tool to weigh individual galaxies and probe their radial mass profile~\citep[e.g][]{Warren2003,Wayth2005,Bolton2012}. Large samples of source-selected lenses are now available thanks to the work of multiple teams over the last two decades~\citep[e.g.][]{Kochanek1995,Munoz1998,Browne2003,Myers2003,Oguri2006,Oguri2008,Cabanac2007,Faure2008,More2011,Inada2012,More2016,Williams2017,Agnello18}. Recent studies search for strong lenses through careful processing of large imaging datasets~\citep[e.g][]{Joseph2014,Paraficz2016,Ostrovski2017} or through citizen science projects such as Spacewarps~\citep{SpaceWarps}.

Whereas source-selected samples span a wide range of physical properties of the lenses, lens-selected samples enable us to study specifically the targeted lenses. The largest lens-selected sample available to date is the Sloan Lens ACS Survey \citep[SLACS,][]{Bolton2006,Bolton2008,Auger2009}, where the lenses are early-type galaxies with redshift $0.16<z<0.49$ selected from the Sloan Digital Sky Survey (SDSS). The SLACS survey used spectroscopic data to search for extra emission lines superimposed on the foreground galaxy spectra, following the method of \citet{Warren1996}.

Motivated by the success of SLACS \citep{Bolton2006,Shu2017}, the Optimal Line-of-Sight lens survey \citep[OLS-lens survey,][]{Willis2006}, the BOSS Emission Line Lens Survey \citep[BELLS,][]{Brownstein2012}, the SLACS for The Masses Survey \citep[S4TM,][]{Shu2015}, the BELLS GALaxy-Lyman-$\alpha$ EmitteRs Systems survey (BELLS GALLERY, \citealt{Shu2016,Shu2016b}) and the subsequent confirmation of the majority of its Galaxy-LAE candidates \citep{Cornachione2017}, we decided to explore further the potential role of QSOs as lenses.

In light of these previous works, we believe that both lens- and source-selected strong lenses samples can now be obtained from wide-field spectroscopic surveys, when the foreground object is easily identified. Selection of the lensed object is enabled by the detection of specific emission lines such as Lyman-$\alpha$ or the presence of both [OII]$\lambda\, 3727$ \AA, H$\beta$, [OIII]$\lambda\, 4959$ \AA, [OIII]$\lambda\, 5007$\AA\, and/or H$\alpha$ suggesting the presence of higher redshift Lyman-$\alpha$ Emitters (LAE) and emission-line galaxies, respectively.

\citet{Miralda1992} have argued that galaxy-galaxy strong lenses should be ubiquitous, and the above surveys have overcome the challenge of their detection. However, it is not the same for QSO-galaxy lenses. In a pioneering study, \citealt{Courbin2010} have identified 14 QSOs acting as potential strong gravitational lenses and confirmed $3$ \citep{Courbin2012} using the \textit{Hubble Space Telescope (HST)} deep imaging capacities (program GO\#12233, Wide Field Camera 3 and UVIS detector). The research was conducted over SDSS-II Data Release 7, fitting and subtracting a spline QSO continuum before cross-correlating the residuals with appropriate emission line templates. However, the small number of confirmed QSO lenses limits the analysis to the intrinsic properties of the individual targets. The promise of a statistically significant sample of QSO lenses is to compare their dynamical and lensing mass distribution and test the scaling laws between the QSO emission lines, the black hole mass and the host galaxy total mass (e.g. \citealt{Kaspi2005}, \citealt{Shen2008}). In a more recent study, \citet{Cen2017} have shown that strong lensing by QSOs could act as an efficient test of different models of dark matter halos of QSO host galaxies \citep{Shen2013,Cen2015}. With our new sample, we can increase the number of known QSO lenses by up to a factor of 3 or 4 and open the door to streamlined detection of such objects in future wide-field surveys. This would in turn benefit studies linking QSOs to their host galaxies.

In this paper, we present 12 new QSOs potentially acting as strong lenses in the SDSS-III BOSS Data Release $12$ \citep[DR12][]{SDSS,BOSS,DR12,Smee2013} as well as our selection method which extends spectroscopic selection of strong lenses to foreground QSOs. In Section \ref{section_method}, we review our candidate selection method and provide spectroscopic evidence for all $12$ candidate systems, as well as photometric hints for probable lensing features for one QSO-ELG lens candidate. In Section \ref{section_qsogal}, we show the evidence for the first $9$ candidates being probable QSO-ELG strong lensing systems presenting higher redshift OII, H$\beta$ and OIII on top of their QSO spectrum. The remaining 3 candidates in Section \ref{section_qsolae} present an asymmetric single line emission feature in their spectra that cannot be attributed to the QSOs emission lines, indicating possibly a high-redshift LAE lensed by the QSO. In Section \ref{section_discussion} we discuss the completeness of our algorithm search for quasar lenses. We then review the number of QSO-ELG candidates obtained compared to \citet{Courbin2010}.

\begin{figure*}
\centering
\includegraphics[width = 0.9\textwidth]{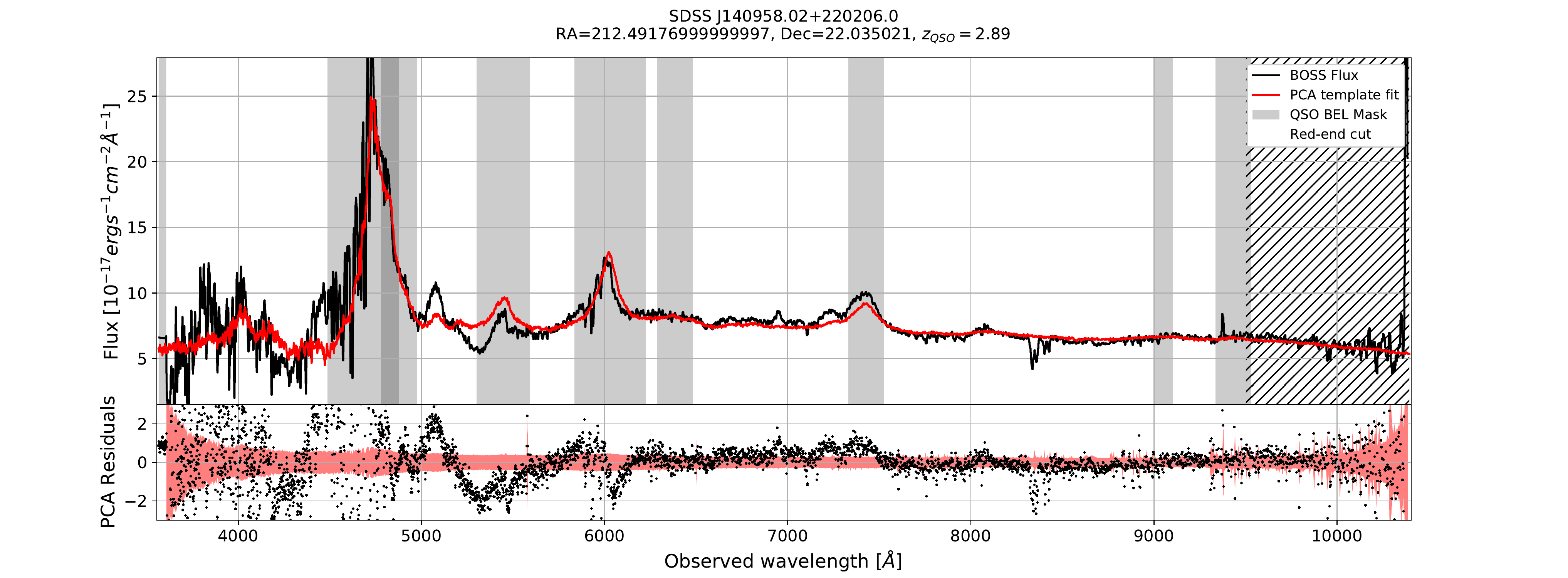} 
\caption{Typical example of a large misfit between the SDSS pipeline PCA template and the QSO spectra.  The gray shaded regions are the masked QSO emission line regions of Table \ref{mask_table_QSO} and sky emission lines of Table \ref{mask_table_sky}. The measured flux is in black, the pipeline PCA template spectrum in red. The lower panel shows the residuals from the PCA template fit subtraction in black and the SDSS spectrograph $1\sigma$ error array in red. Note the large residuals, even outside masked regions which create many false positives when searching for extra emission lines from a different object in the QSO spectrum. }
\label{mask_fit}
\end{figure*}

\begin{figure*}
\centering
\includegraphics[width = 0.48\textwidth]{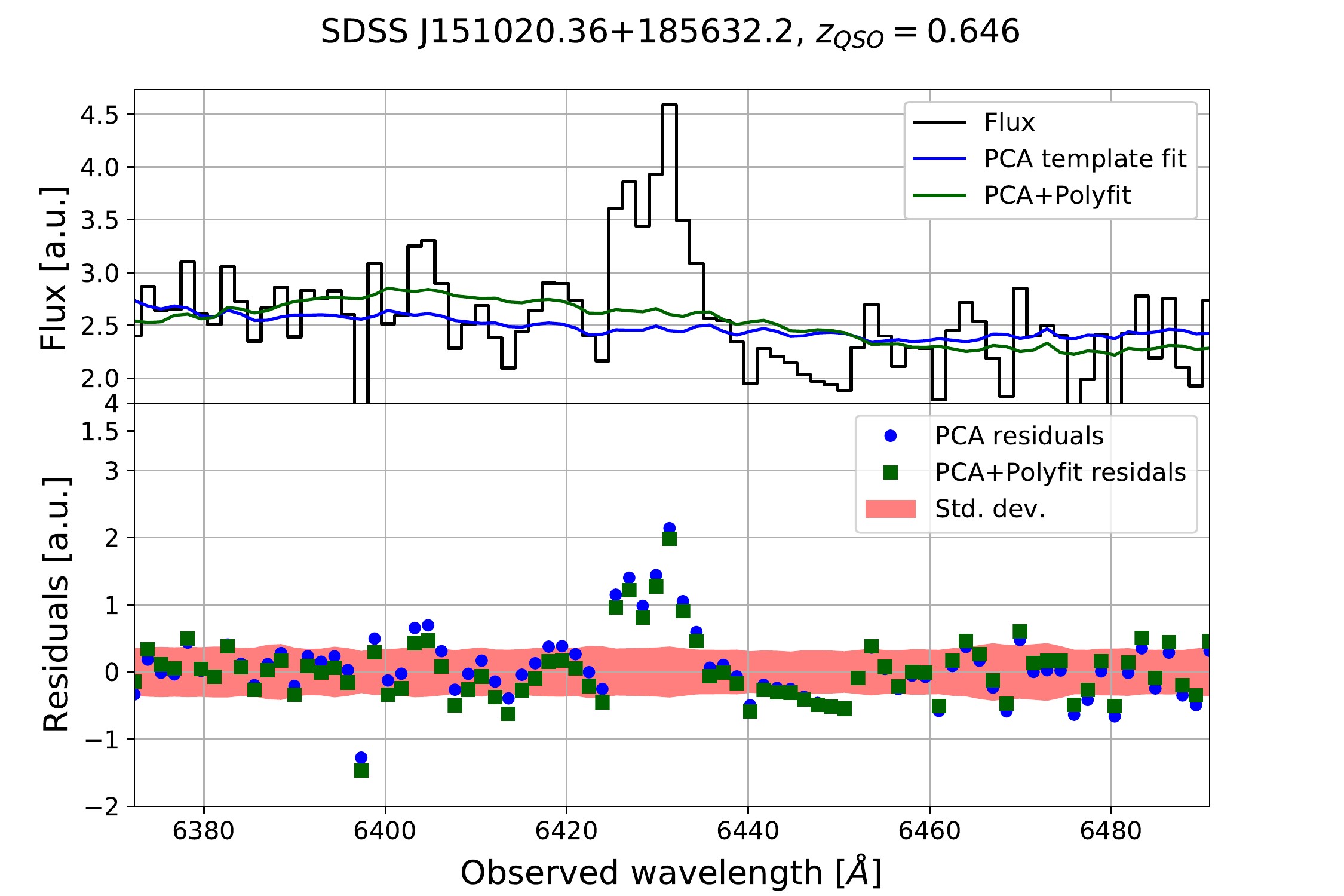} 
\includegraphics[width = 0.48\textwidth]{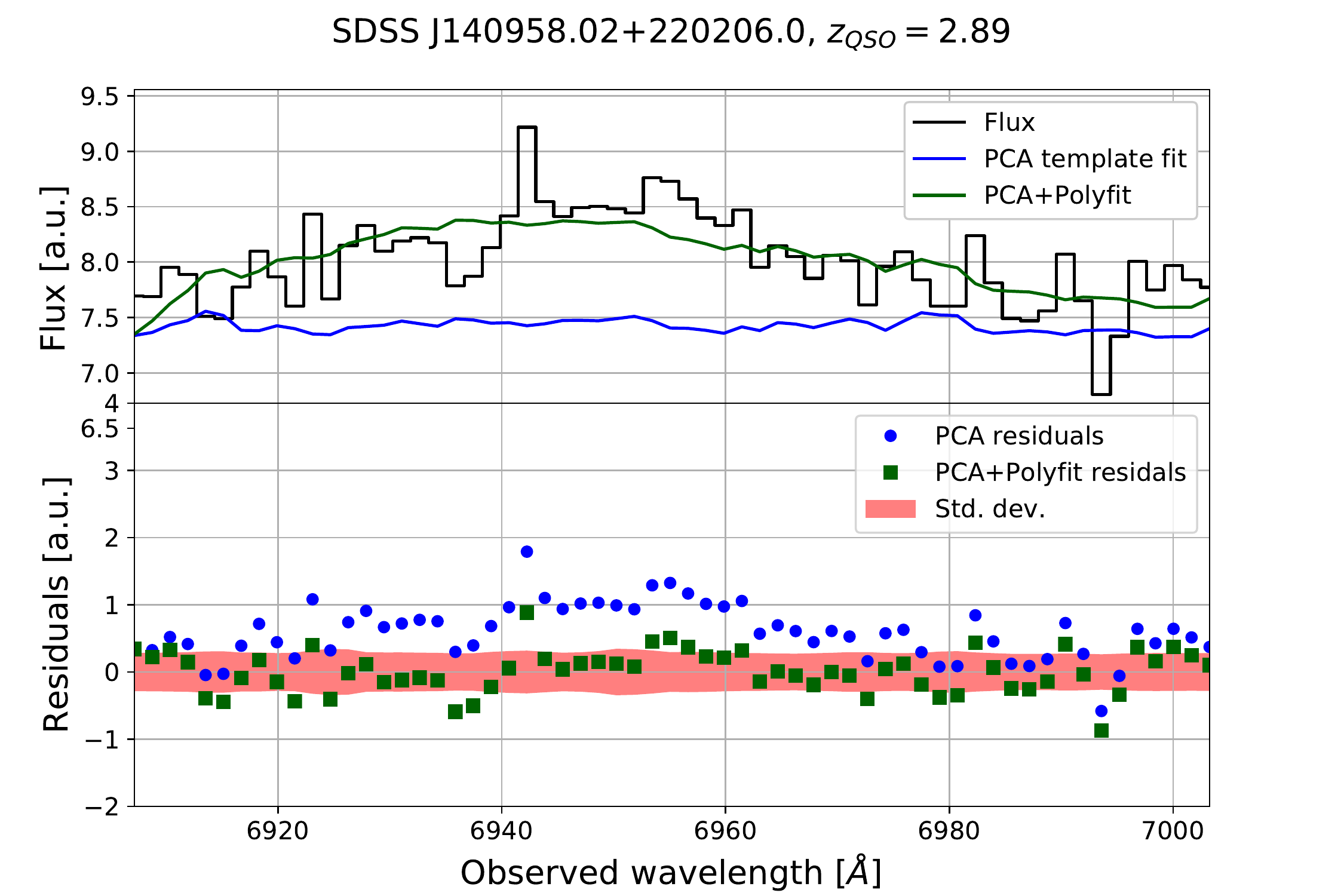} 
\caption{Comparison between a true emission line detection (right) and a false detection generated by an incorrect PCA template fit (left). However, locally fitting (and subtracting) a third order polynomial to a $40$ \AA \, section of the spectra around the line candidate is sufficient to remove continuum features unsubtracted by the PCA. The flux is in black, the PCA template in blue, and the polynomial fit in green. The lower panel showcases the SDSS spectrograph $1\sigma$ error array in red, and the residuals from both fits in the same color code. Note the residuals after the subtraction of the polynomial fit. The residuals of the PCA fit are removed and do not result in a detection anymore, whereas the thin emission line is still clearly detected.}
\label{fig:polyfit}
\end{figure*}

\section{Mining BOSS for candidate QSO lenses}
\label{section_method}
The SDSS-III BOSS survey provides an unparalleled sample of $297301$ QSO optical spectra, with wavelength coverage $3600\,$\AA \, to $10400\,$\AA \, and resolution $R\approx 1500-2000$, all inspected and confirmed by eye in \citet{Paris2017} for Data Release 12. This datasets yields a huge potential for the discovery of QSOs acting as strong lenses. In this section, we detail our search technique.

The selection method used for our sample is based on the one used for the BELLS, SLACS, S4TM and BELLS GALLERY surveys, but with significant changes to select Lyman-$\alpha$ or nebular emission lines superimposed on QSO spectra. The basic principle is to identify additional emission lines distinct from the expected spectral features of the foreground QSO. In this study, we first search for QSO-Emission Line Galaxy (ELG) systems by looking for additional OII, H$\beta$, OIII and/or H$\alpha$ lines, limiting the redshift of the source to $z_s\lesssim 0.9$ due to the spectrograph wavelength range. Secondly, we search for QSO-Lyman-$\alpha$ Emitter (LAE) systems, where the asymmetry of the Lyman-$\alpha$ is a signpost for the observed line, allowing us to derive the redshift of the source even though only one line is detected. Detecting Lyman-$\alpha$ emission from a background source in SDSS spectra implies $2 \lesssim z_s\lesssim 6.8$. The lower limit for any background source redshift is also constrained by the foreground QSO redshift.

Detecting QSOs (rather than galaxies) acting as strong lenses \textit{via} the detection of additional lines is difficult for several reasons. The SDSS pipeline fits the QSOs with a template constructed from the first five eigenvectors of Principal Component Analysis (PCA) decomposition performed on a chosen set of SDSS DR7 QSOs \citep{Paris2011}, which provides correctly estimated redshifts for most of DR12 QSOs. However, the PCA template is far from perfect and is known to have serious limitations when it comes to reproducing strong broad emission lines or less frequent lines in QSOs spectra (see Fig. 1). Even though we can take advantage of the PCA to approximate the QSO spectra at first order, a range of features are bound to missed by the PCA and might be mistaken for background  features. By contrast, the spectral templates of galaxies perform well for SDSS galaxies for that purpose \citep[e.g.][]{Bolton2006,Brownstein2012,Shu2016}. 
This issue can be overcome by masking broad emission lines where the PCA is most likely to fail, but this limits the search range, and a time-consuming visual inspection and selection of lens candidates is still necessary to ensure a robust selection. Last but not least, the QSO outshines the lensed background source. This prevents the confirmation of the lensing systems with wide-field photometric data and highlights the crucial importance of high-resolution imaging to confirm such lenses\citep{Courbin2010,Courbin2012}.

\subsection{Preparation of the dataset}
Given the limitations of the PCA for the task at hand, we select a subset of the BOSS DR12Q \citep{Paris2017} dataset to limit future false positive detections. This procedure is summarized in these following steps. 
\begin{enumerate}
\item We retain only spectra classified as 'QSO' by the SDSS pipeline and remove any QSO found in the blank sky fibers. We discard any QSO fitted by the pipeline with flags \verb+ZWARNING+$!=0$, \verb+Z_ERR+$<0$ or a reduced $\chi^2_{fit}>10$, which signal potentially wrong redshifts or poor continuum fitting. We also drop any QSO that has a PCA redshift differing by $>\Delta z>0.1$ from its visually determined redshift. 
\item The best-fitted QSO PCA template spectrum provided by the BOSS pipeline is subtracted from the spectra. 
\item We mask the main QSO emission lines by setting the inverse variance of the spectra to $0$, preventing any detection of residual QSO emission lines. The same procedure is applied to sky lines to prevent any spurious detection. The masked emission lines are presented in Tables \ref{mask_table_QSO} and \ref{mask_table_sky} as well as graphically in the upper panel of Fig. \ref{mask_fit}. The masked fraction of the observed spectra is typically about $\sim 25\%$. Nonetheless, this masked fraction increases with redshift for multiple reasons. First of all, emission lines cannot be searched in the Lyman-$\alpha$ forest of the quasar, reducing dramatically the search space at $z\gtrsim 2$. Secondly, rest-frame QSO UV broad lines (e.g. N{~\small V},Si{~\small IV}, C{~\small IV}, Mg{~\small II}) are typically broader than rest-frame optical lines. Finally, the observed width of all lines increases with redshift. We detail in Section \ref{sec:completeness} the impact of the masking on the completeness of the lens search.
\end{enumerate}

\subsection{First emission feature search}
After subtracting the PCA template from every suitable QSO spectra selected above, we search for high-Signal-to-Noise (SN) features in the residuals. In order to do this, we make use of a simple matched-filtering approach which is a key step of all spectroscopic selection methods used in the SLACS, BELLS, S4TM and BELLS GALLERY strong lensing samples and was first presented in Sect. 3.1 of \citet{Bolton2004}. We reproduce it here for self-consistency.

The matched-filter search done by convolving a Gaussian kernel $\{u_i\}$ with the spectrum residuals $\tilde f_i$. The maximum likelihood estimator of the line flux at pixel $j$ with amplitude $A_j$ is the one minimizing the $\chi^2$ value
$$\chi^2_j = \sum_i \frac{\left( A_ju_i -\tilde f_{(j+i)}\right)^2}{\sigma^2_{(j+i)}} \quad , $$
where $\tilde f_i$ is the reduced flux at bin $i$ and $\sigma_i$ the measured variance. Setting the derivative by $A_j$ to $0$ gives the maximum likelihood estimator 
$$\overline{A}_j  = C_j^{(1)} / C_j^{(2)} \quad , $$
where the two coefficients $ C_j^{(1)}$,  $C_j^{(2)}$ are defined as 
$$ C_j^{(1)} = \sum_i \frac{\tilde f_{j+i} u_i}{\sigma_{(j+i)}^2} \quad , $$
$$ C_j^{(2)} = \sum_i \frac{u_i^2}{\sigma_{(j+i)}^2} \quad .$$
Assuming uncorrelated Gaussian errors on $f_i$, the SN of the estimator $\overline{A}_j$ is 
$$SN_j = C_j^{(1)} / \sqrt{C_j^{(2)}} \quad .$$

This estimator is quick to compute on all pixels of the residuals, giving an estimate of the SN of potential Gaussian emission features at each pixel. In the following, we refer to this estimator as the SN for simplicity. The matched-filter feature search is performed on all pixels of each spectrum. We use a Gaussian kernel with a dispersion $\sigma = 150 \text{km}\text{s}^{-1}$ and store detections above $8\sigma$.

\subsection{Algorithmic search for background features}
\label{section_algo}
In this section, we describe our algorithm for selecting secure background objects based on the detections of significant features described above.
We first search for multiple nebular emission lines at a similar redshift present in the QSO spectrum residuals to identify QSO-ELG lens candidates. Then, we sort the remaining extra single emission lines to select QSO-LAE lens candidates using the distinct asymmetry of the Lyman-$\alpha$ profile.

\begin{enumerate}
\item  We limit this search to the region $(1+z_{QSO})\lambda_{Lyman-\alpha}<\lambda_{\text{observed}}<9500 \text{ \AA}$ to avoid the Lyman-$\alpha$ forest and the red edge of the SDSS spectrograph, highly populated by sky emission lines. All matched-filtered detections outside this range are immediately discarded.
\item A 3rd order polynomial is fitted in a $\sim 6000 \text{km s}^{-1}$ velocity range centered on detections to remove any contribution from a poor PCA continuum fit. The SN is recomputed and we keep only features with a new SN above $6\sigma$ to account for the fact that the polynomial might cause a small SN loss to true positives. Large unsubtracted continuum residuals will be discarded at this step, as the SN will drop well below the required $6\sigma$. Figure \ref{fig:polyfit} illustrates this process on both a typical true- and false-positive detections.
\item For each QSO, the detection candidates are ranked by SN and only the five highest are kept. Multiple background sources in the 2'' SDSS fiber are extremely unlikely to be detected. In practice, quasars with more than $5$ lens candidates have an extremely poor continuum fit or present a lot of noise in their spectra. If all background optical emission lines ([OII], H$\beta$, [OIII] and H$\alpha$) were strong enough to be detected with the gaussian-matched filter, we would expect at most $5$ detections. We thus limit the number of candidates to $5$ to reduce the number of candidates that will need to be visually inspected.
\item We first check if each detection is possibly part of a larger group of lines at a higher redshift than the QSO as it is unambiguous signal for an aligned ELG. We ascribe each peak in turn to OII, H$\beta$, OIII and H$\alpha$ at $z_{ELG}\neq z_{QSO}$. We then look up the SN at the expected position of the $4$ other redshifted emission lines. If the quadrature sum SN of the pixels at the expected line positions is greater than initial detection SN by at least $2.5\sigma$ (for a total SN of 10.5$\sigma$), we flag this candidate as a 'QSO-ELG' lens candidate. A total of $254$ 'QSO-ELG' objects were flagged.
\item This leaves $929$ detections that cannot be safely attributed to an ELG and are hence dubbed 'QSO-Single Line Emitter' system (QSO-SLE). They are subsequently visually inspected for asymmetry in the detected line as a hint for Lyman-$\alpha$ emission.
\begin{figure*}[t]
\centering
\begin{tabular}{p{0.5\textwidth} p{0.5\textwidth}}
\includegraphics[width = 0.49\textwidth]{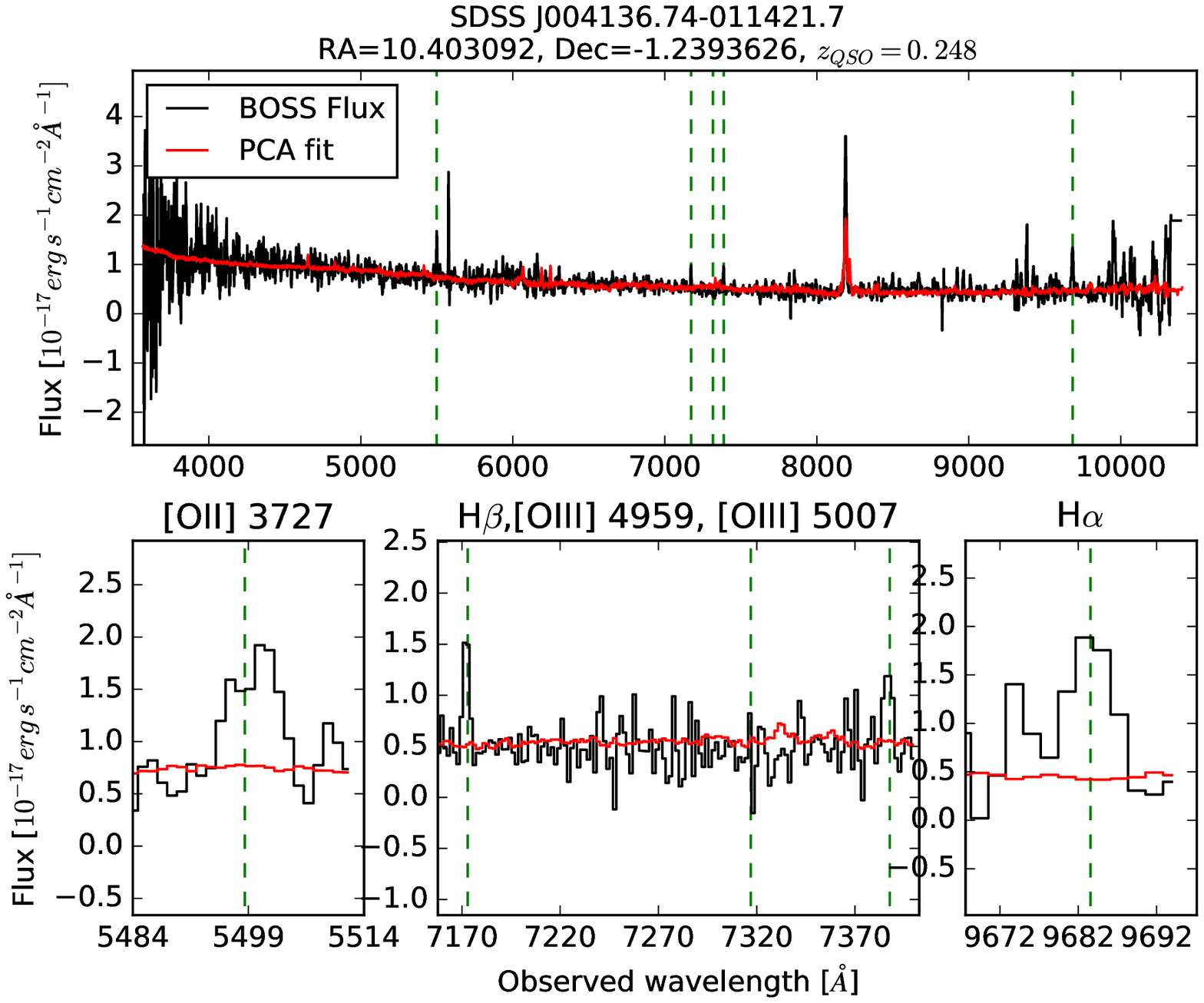} &
\includegraphics[width = 0.49\textwidth]{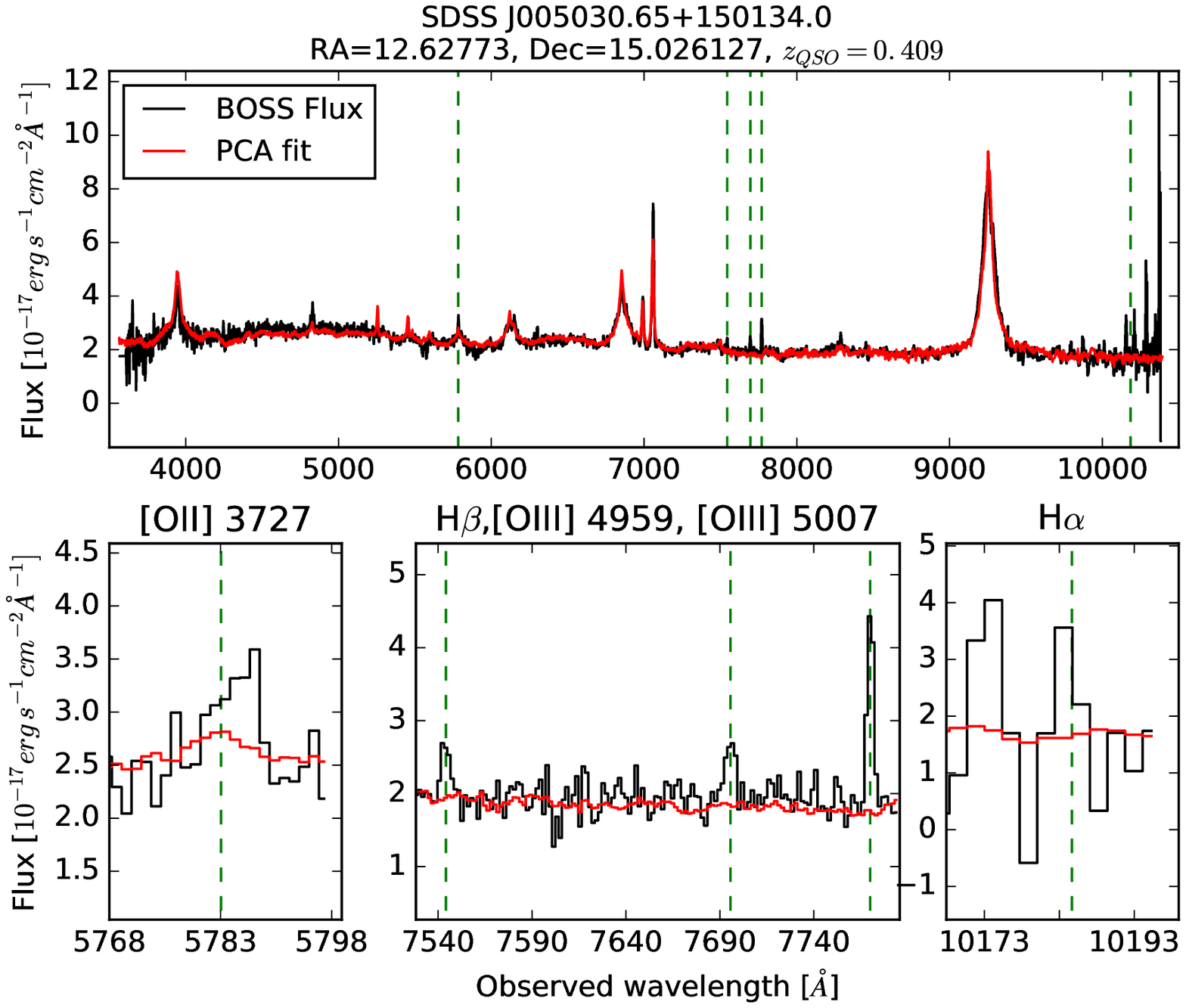} \\
\includegraphics[width = 0.49\textwidth]{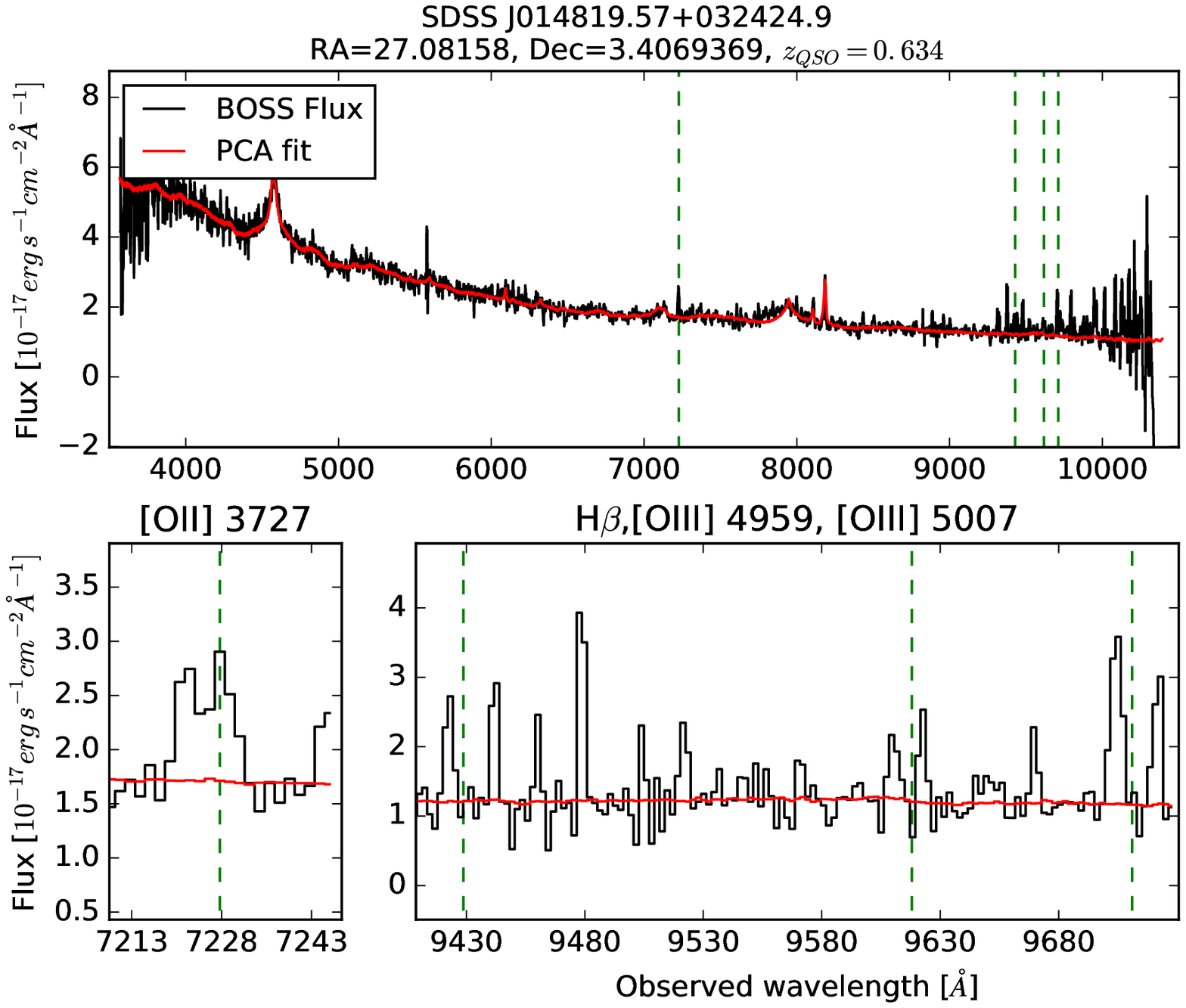} &
\includegraphics[width = 0.49\textwidth]{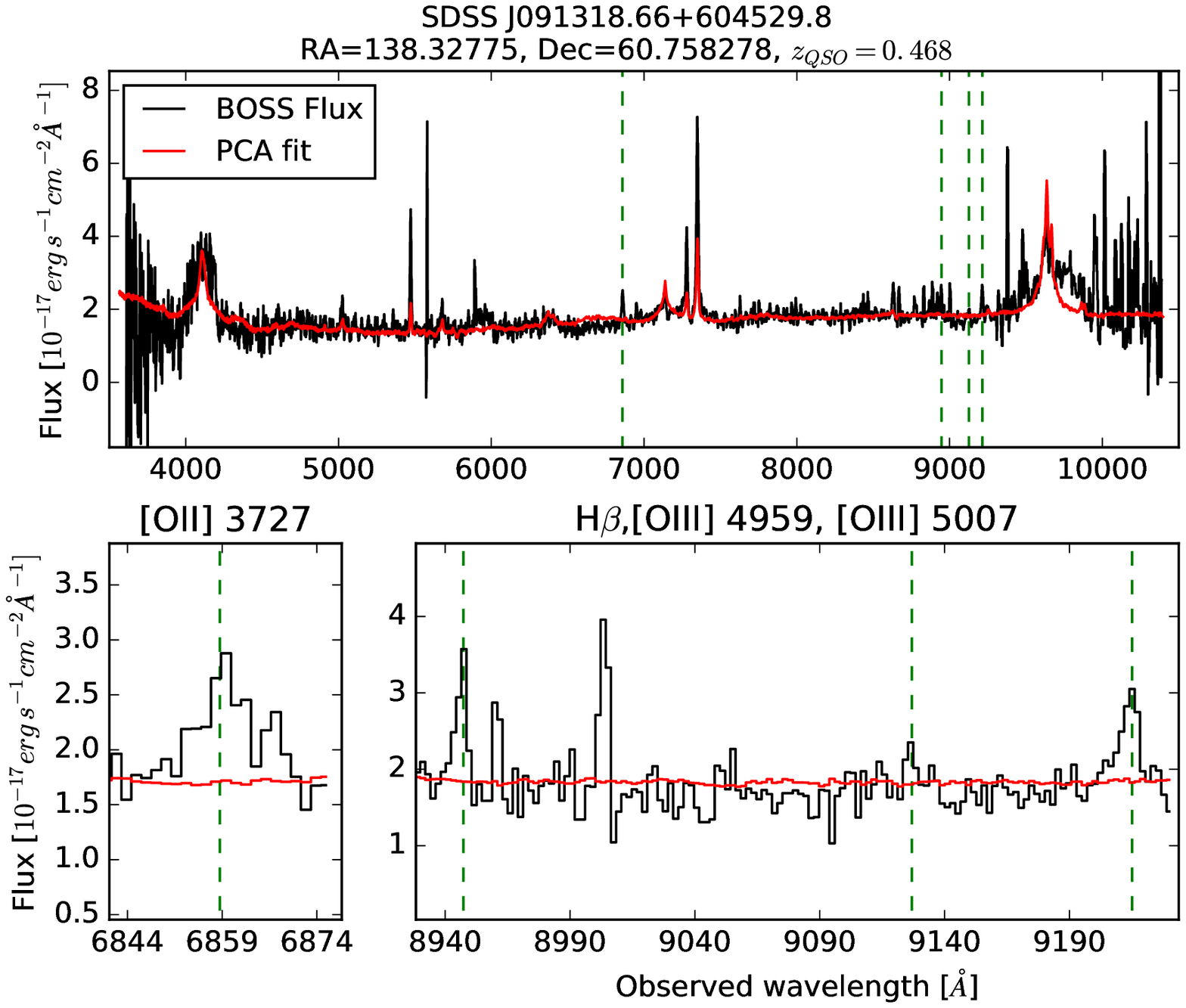} \\
\includegraphics[width = 0.49\textwidth]{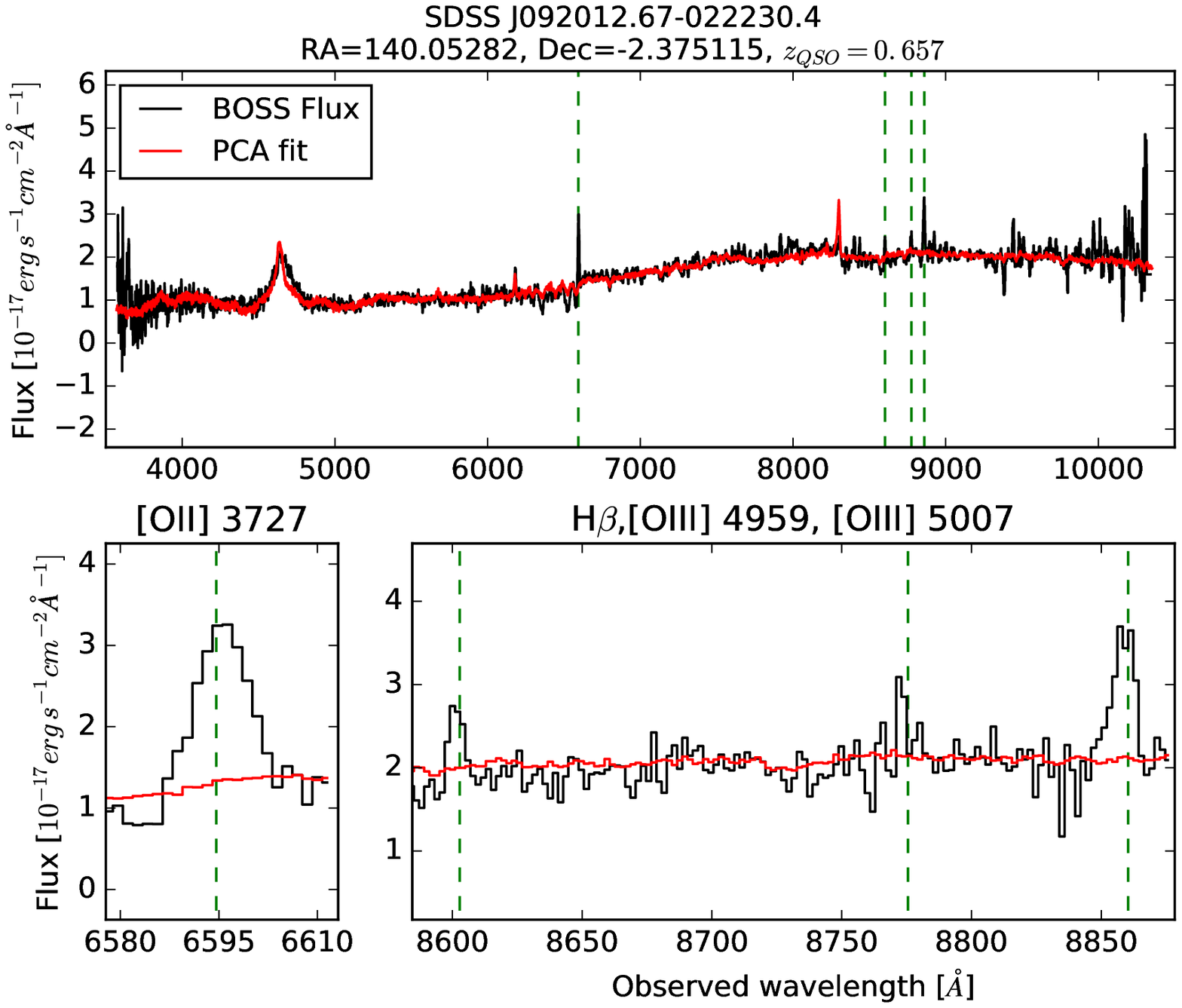} &
\includegraphics[width = 0.49\textwidth]{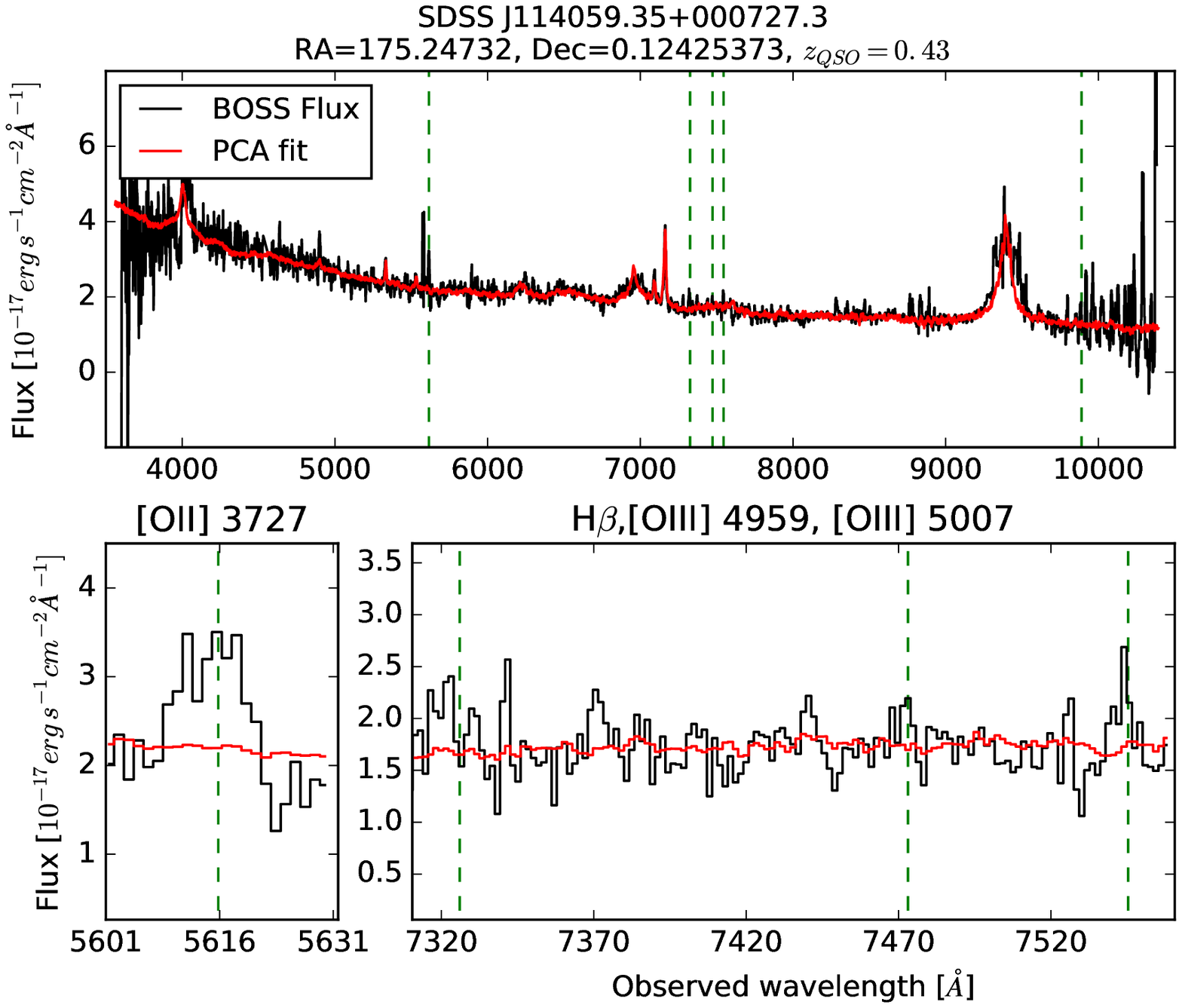}
\end{tabular}
\caption{The 6 first QSO-ELG candidates. Upper panels: 5-pixel smoothed BOSS spectra (black) and QSO fitted PCA template (red). The location of the background galaxy emission lines are indicated by dashed green lines. Lower panels: Zoom on the detected emission lines of the source galaxies.}
\label{QSOGal_candidates_plot}
\end{figure*}
\end{enumerate}

Eventually all QSO-ELG and QSO-LAE lens candidates are inspected by eye and their SDSS or Dark Energy Spectroscopic Instrument
(DESI) Legacy Imaging Surveys \citep[DELS, ][]{Dey2018} imaging is checked to remove any hits possibly due to nearby galaxies. The nearby fibers are also inspected to check for any strong line at the detected wavelength to avoid any false positive due to cross-talk between neighbouring fibers. This leaves a total of $9$ secure QSO-ELG candidates presented in Table \ref{QSOGal_best}. Most of the removed algorithmic candidates were obvious failures of the template and polynomial fits, or unconvincing lines in noisy parts of the spectra. Through visual inspection, we also selected $3$ potential QSO-LAE candidates presented in Table \ref{QSOLAE_candidates_BOSS}. We also record $49$ Single Line Emitter (SLE) detections that were not discarded during visual inspection but were not deemed sufficently asymetric to be attributed to a background LAE (see Appendix \ref{appendix:SLE}). The SLEs redshift distribution is presented in Fig. \ref{histo} alongside the QSO-ELG and QSO-LAE candidates source and lens redshift distributions. 

\section{Lens candidates samples}
\subsection{QSO-ELG candidates }
\label{section_qsogal}
\begin{figure*}
\centering
\includegraphics[width = 0.49\textwidth]{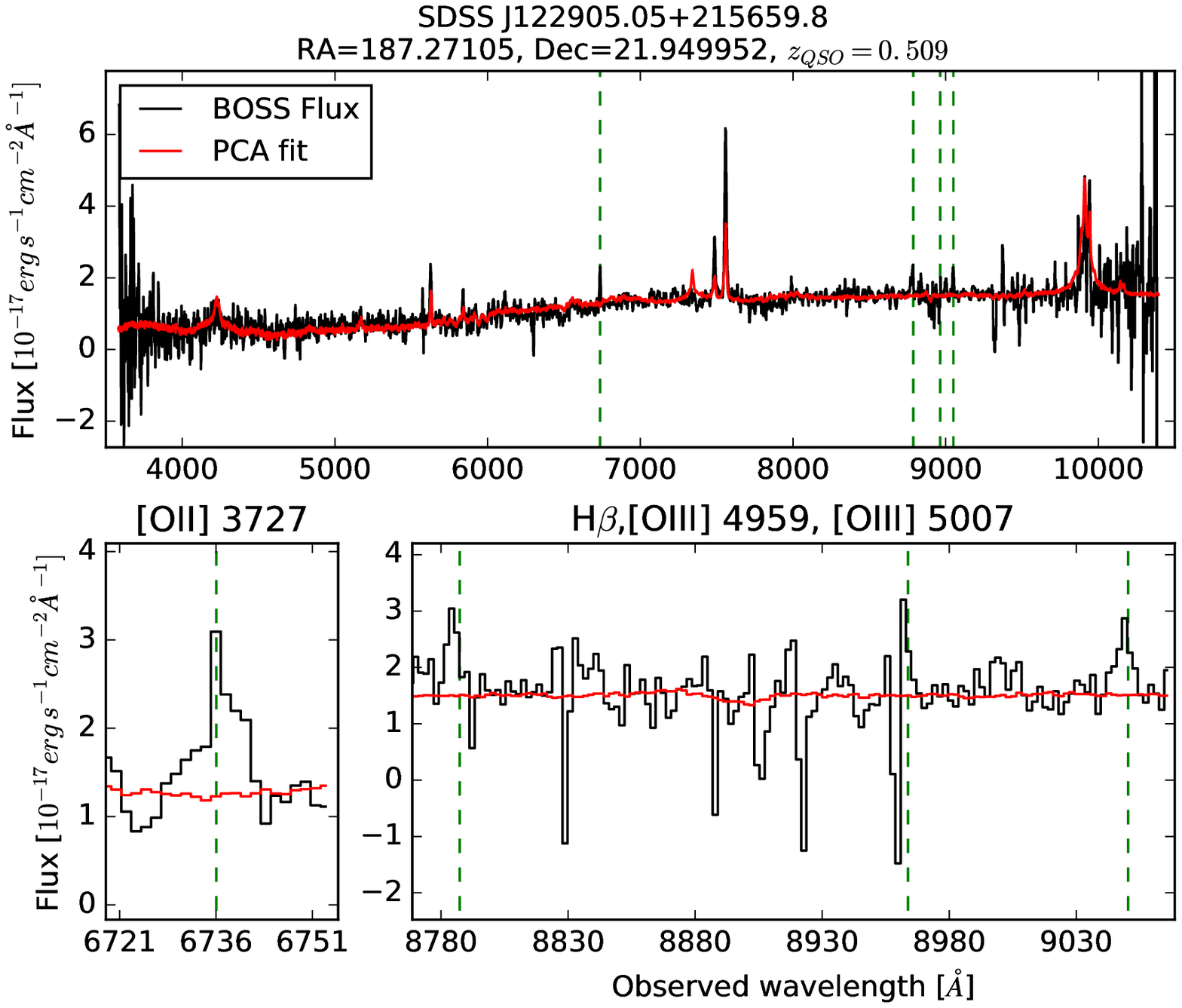} 
\includegraphics[width = 0.49\textwidth]{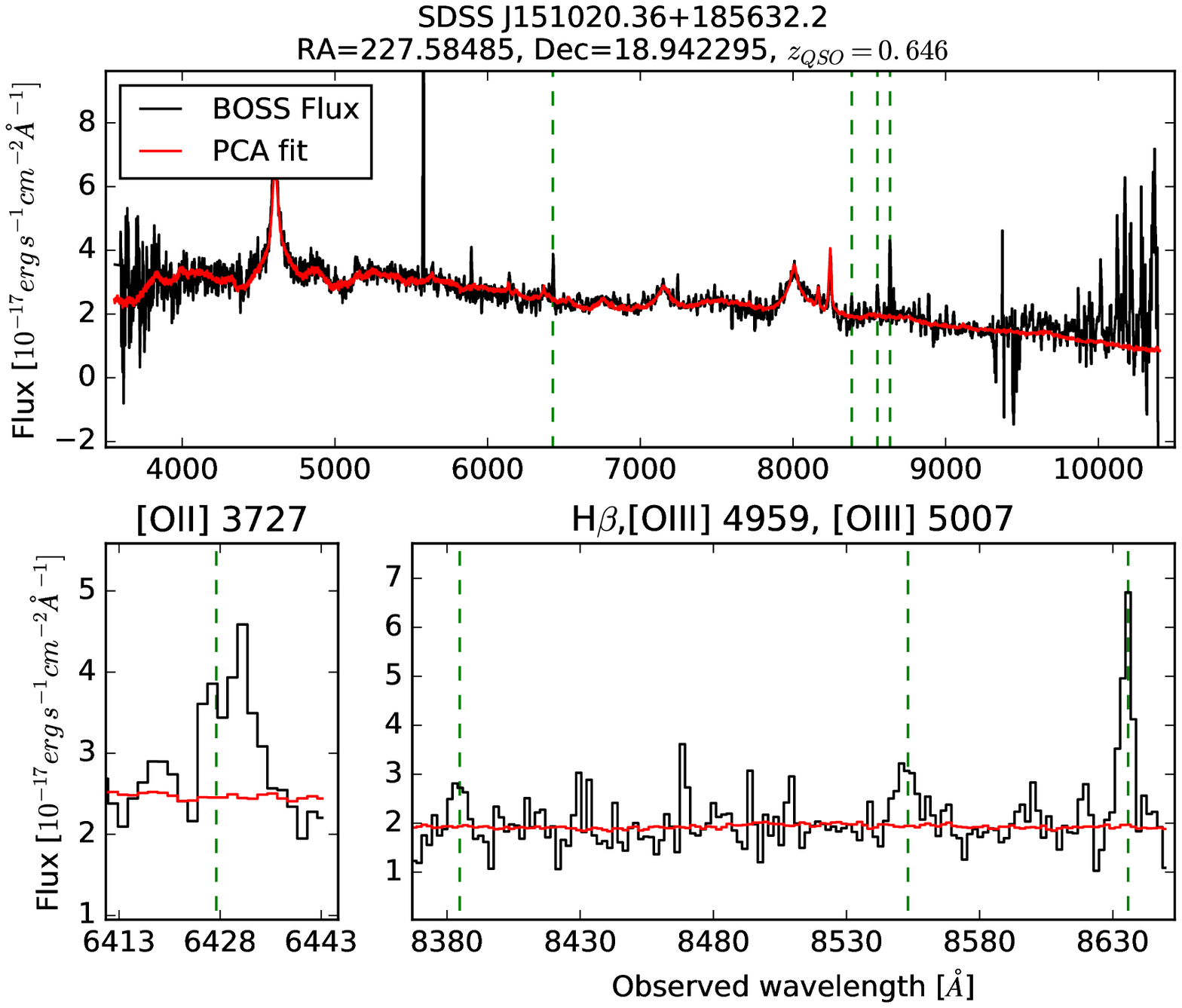} \\
\includegraphics[width = 0.49\textwidth]{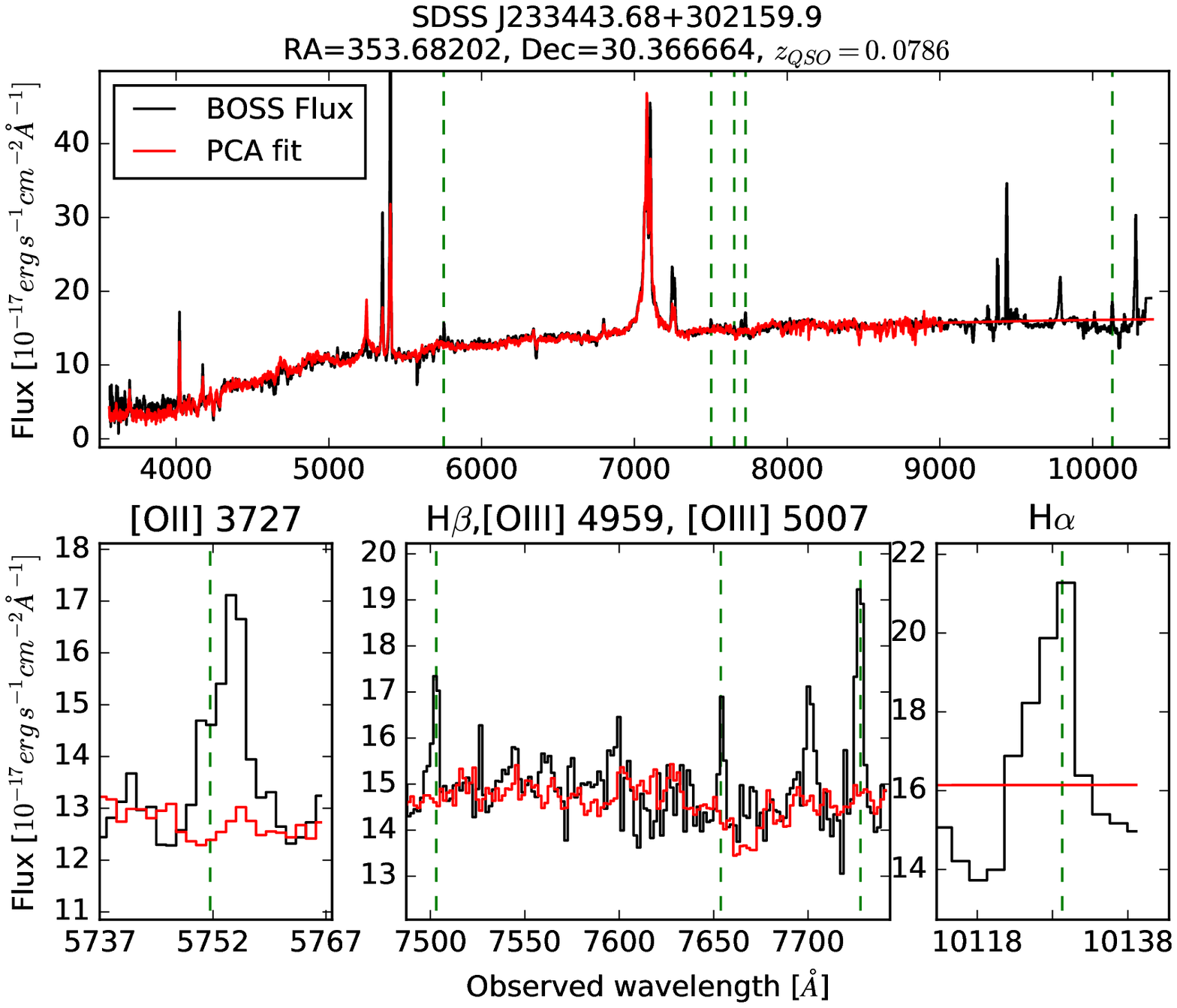} 
\caption{Spectroscopy of the 3 last QSO-ELG system candidates. Upper panels: 5-pixel smoothed BOSS spectra (black) and QSO fitted PCA template (red). The location of the background galaxy emission lines are indicated by dashed green lines. Lower panels: Zoom on the detected emission lines of the source galaxy.}
\label{QSOGal_candidates_plot_2}
\end{figure*}

\begin{figure}
\centering
\includegraphics[width =\textwidth]{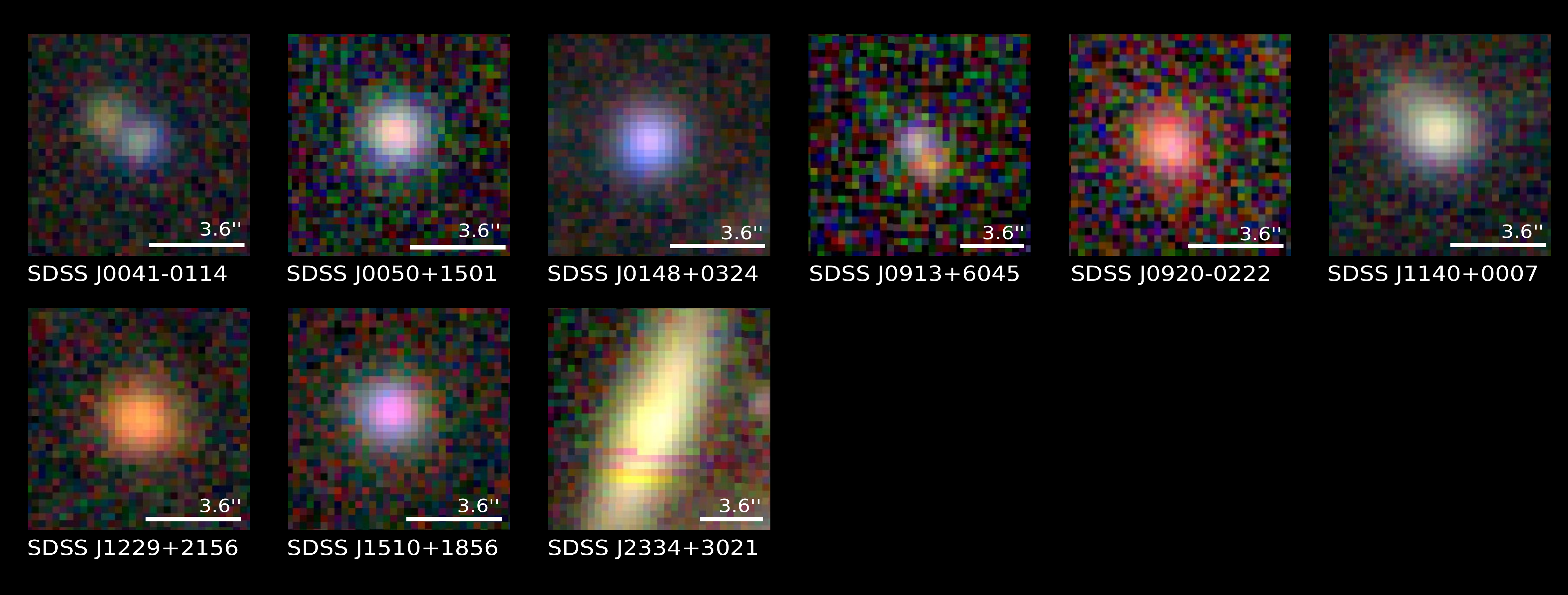}
\caption{DELS / SDSS imaging of the $9$ QSO-ELG candidates. The stamps are centered on the BOSS targets, and the thick white line indicates the diameter of the BOSS fiber ($2''$) augmented by the mean $80$th percentile seeing ($\sim 1.6''$). As expected, no candidates prevent evidence of lensed features, but nearly half of them display a nearby redder object.}
\label{phot_candidates_gal}
\end{figure}

The $9$ QSO-ELG candidates were selected to present clear visual evidence of extra background emission lines. All of them are selected using the above $8\sigma$ detection threshold of an initial line (usually OII or $H\beta$), along with evidence for other lines by requiring a  quadrature sum SN of all lines $>10.5 \sigma$. In practice, all but one have a total SN $>10.5\sigma$. The $9$ candidates all present 4 or 5 significant emission lines among [OII] 3727, H$\beta$, [OIII] 4959, [OIII] 5007 and H$\alpha$ (see Fig. \ref{QSOGal_candidates_plot} and \ref{QSOGal_candidates_plot_2}). Selected properties of the QSO-ELG lens candidate sample are presented in Table \ref{QSOGal_best}, and the source and lens redshift distributions of the systems are presented in Fig. \ref{histo}. 
Three candidates (SDSS J1140+0007, SDSS J0041+0114 and SDSS J0913+6045) present either nearby features surrounding the QSO or a distinct nearby object in the DELS or SDSS color composite images shown in Fig. \ref{phot_candidates_gal}. 

\begin{figure}

\includegraphics[width =0.2\textwidth]{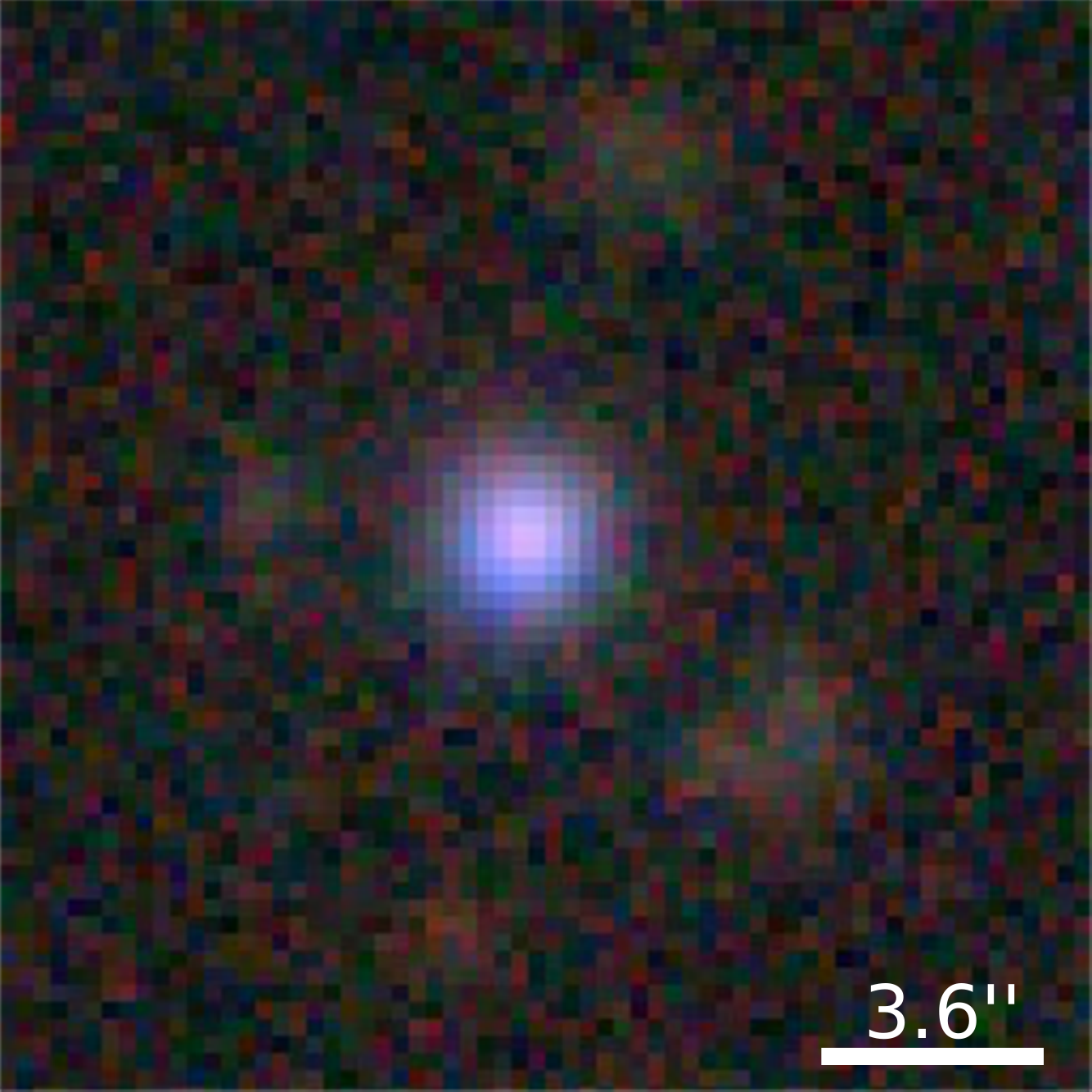}
\centering
\caption{Enlarged view of QSO-ELG candidate J0148+0324. The DELS composite shows four similar faint images in an Einstein cross configuration, albeit at a somewhat large image separation. It is the only candidate featuring such intriguing hints of strong lensing. Surprisingly, this candidate is also remarkable by the presence of not one, but two series of detected background emission lines at a very close velocity separation (see Fig \ref{QSOGal_candidates_plot}). }
\label{fig:J0148_large}
\end{figure}

J0148+0324 presents four distinct features disposed around it in a seemingly Einstein Cross configuration  (see Fig. \ref{fig:J0148_large}). However, we note that the Einstein radius $\sim 4''$ would be quite large for such a strong lensing system. Coincidentally or not, J0148 also presents two close series background emission lines (Fig. \ref{QSOGal_candidates_plot}) at $\sim 400 \text{km}\text{s}^{-1}$ separation. We hypothesize that these intriguing characteristics could be due to a small background  cluster or group lensed by the quasar.

\subsection{QSO-LAE candidate lenses}
\label{section_qsolae}
\label{Lya_section}

\begin{figure*}
\centering
\includegraphics[width = 0.95\textwidth]{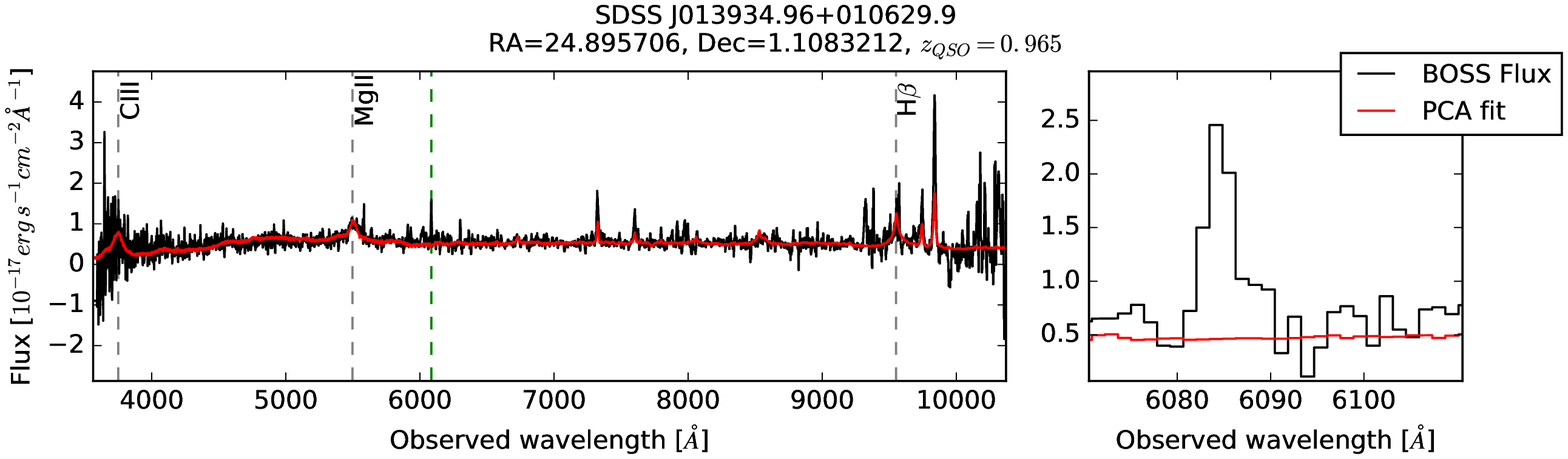} \\ 
\includegraphics[width = 0.95\textwidth]{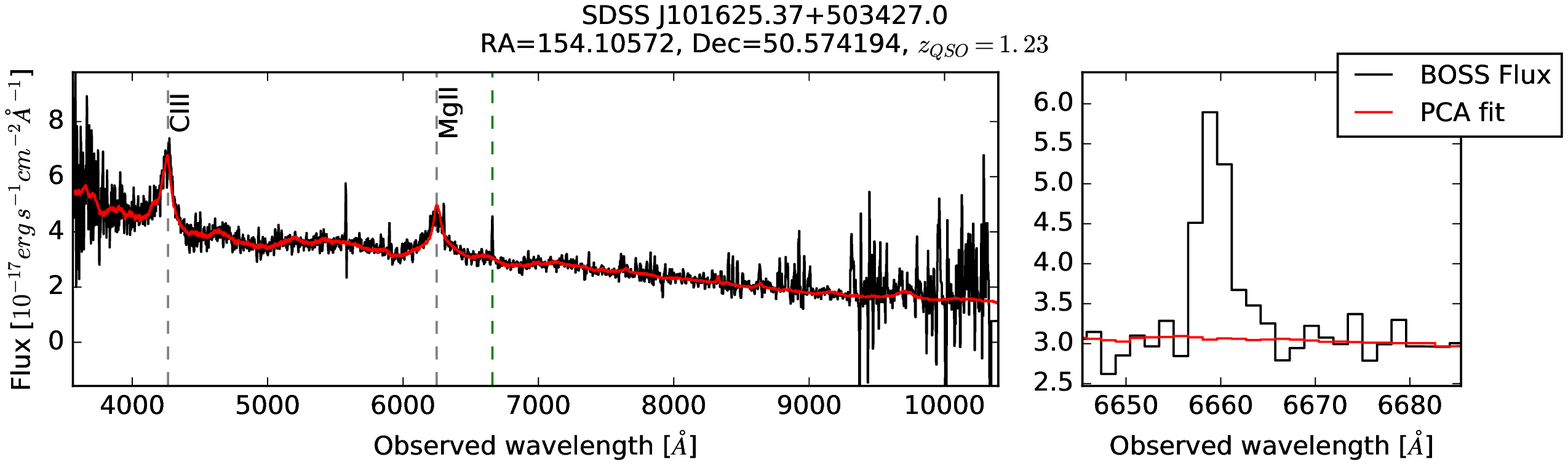} \\ 
\includegraphics[width = 0.95\textwidth]{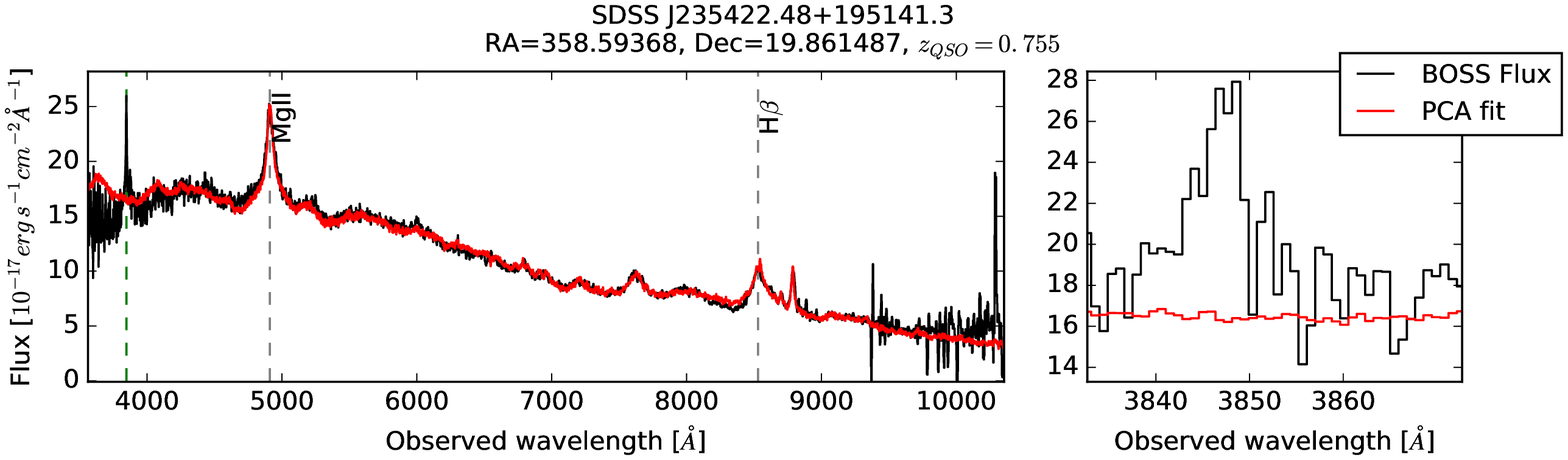}
\caption{SDSS spectra of the QSO-LAE lensing systems candidates. Right panels: 5-pixel smoothed BOSS spectra (black) and QSO fitted PCA template (red). The detected extra emission line is indicated by the dashed green line. Left panels: Zoom on the original spectra of the candidate lensed Lyman-$\alpha$ emission line. Note the asymmetric profile characteristic of Lyman-$\alpha$.}
\label{BOSS_candidates_plot}
\end{figure*}

\begin{figure}
\centering
\includegraphics[width =0.5\textwidth]{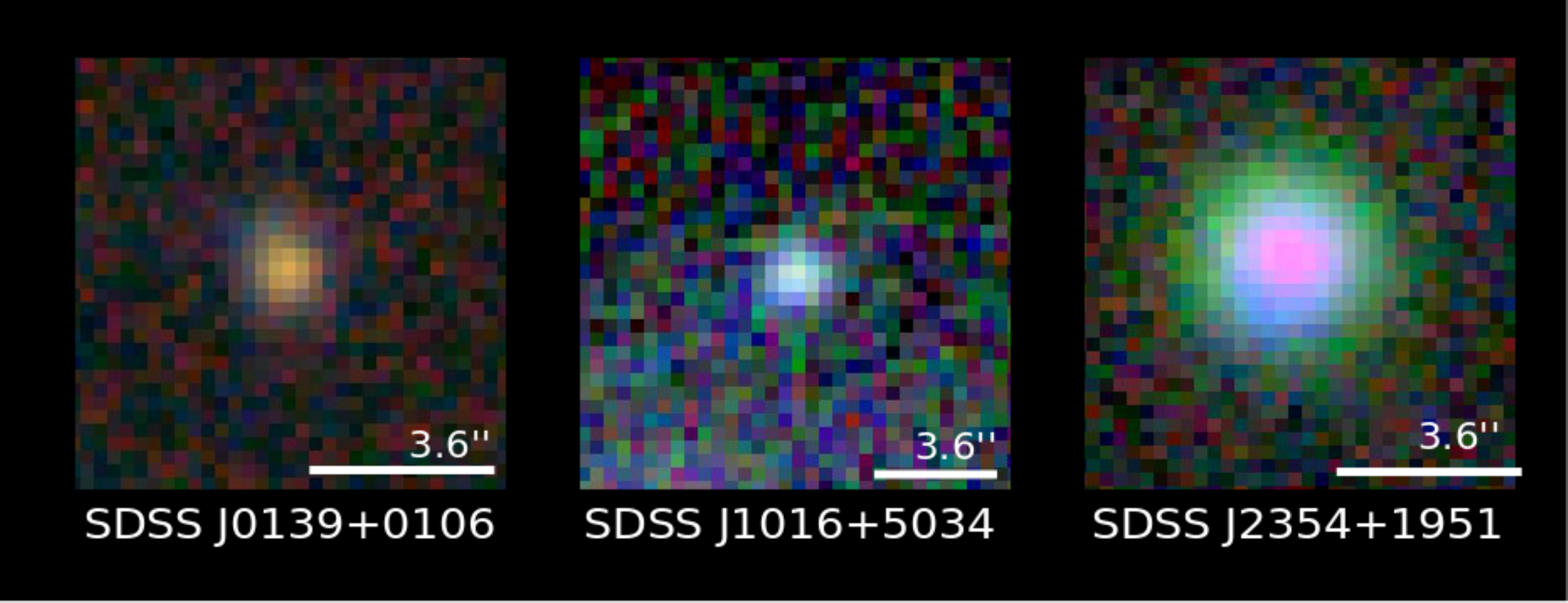}
\caption{DELS / SDSS imaging of the $3$ QSO-LAE candidates. The stamps are centered on the BOSS targets, and the thick white line indicates the diameter of the BOSS fiber ($2''$) augmented by the mean $80$th percentile seeing ($\sim 1.6''$). As expected, no candidates prevent evidence of lensed features.}
\label{phot_candidates_lae}
\end{figure}

\begin{figure}
\hspace{-0.8cm}
\includegraphics[width = 0.5\textwidth]{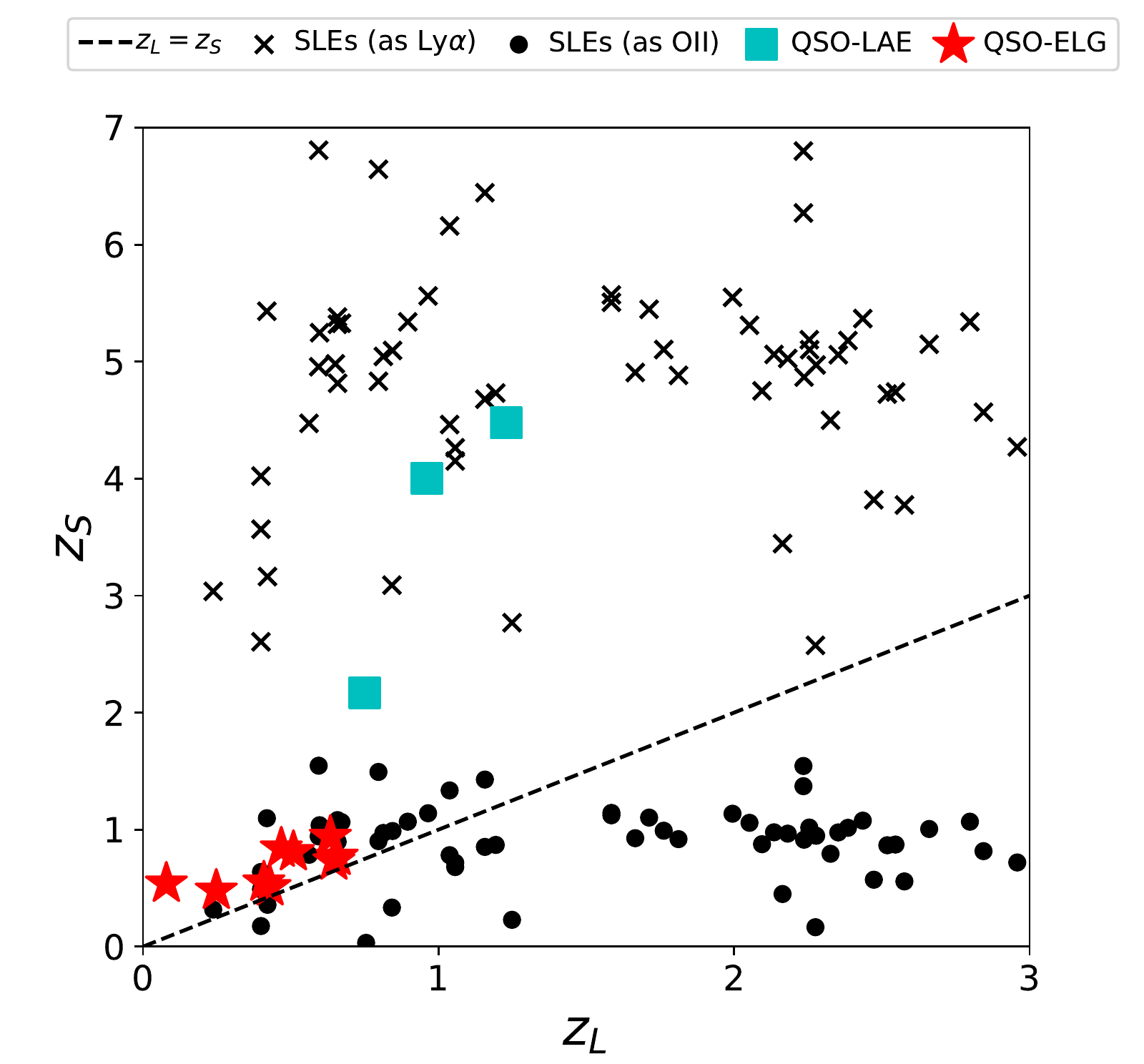} 
\caption{Lens ($z_L$) and source ($z_S$) redshift distributions of the $9$ QSO-ELG lens candidates, the $3$ QSO-LAE lens candidates and the $49$ remaining SLEs selected by the algorithm that were not shortlisted as QSO-LAE candidates (see Appendix \ref{appendix:SLE} for some examples), with redshift are determined assuming either background Lyman-$\alpha$ or [OII] emission. In the case of QSO-LAEs, our selection method seems to favor low redshift QSOs and high-redshift background sources. QSO-ELG candidates however have both low lens and source redshifts. The lower and upper boundaries on the source redshift are due to the limited range of the SDSS spectrograph. The dotted line indicates the $z_S > z_L$ limit for background object detections.
}
\label{histo}
\end{figure}

Our $3$ QSO-LAE candidates were visually selected from the $49$ SLE sample to present asymmetric Lyman-$\alpha$ profiles. We exclude low-redshift OII emission by checking for the absence of emission at the expected redshifted H$\beta$,OIII, $H\alpha$ wavelength. The QSO-LAE candidates redshift distribution is presented in Fig. \ref{histo} along with the other SLEs under the assumption that the single line is either Lyman-$\alpha$ or [OII]. The spectroscopy of the $3$ QSO-LAE is shown in Fig. \ref{BOSS_candidates_plot}. The SDSS or DELS imaging of each QSO-LAE is presented in Fig. \ref{phot_candidates_lae} but does not confirm the presence of strong gravitational lensing features. As in \citet{Courbin2012}, we do expect the lensed images to be outshined by the QSO in the SDSS photometry and thus to be revealed only by high-resolution imaging.

\begin{landscape}
\begin{table}
\caption{Selected properties of the 9 QSO-ELG lens candidate systems}
\label{QSOGal_best}
\centering
\begin{tabular}{lrrcrrrrrrrrrr}
\hline\hline   
Target & RA & DEC & Plate-MJD-Fiber & $z_{QSO}$ & $z_{S}$ & $SN_l$ & $SN_{tot}$ &  $g$  & $i$ &  $M_{SIS}$ & $f_{OII}$ & $f_{OIII}$   \\
\hline
SDSS J004136.74-011421.7 & 10.403092 & -1.2393626 & 4222-55444-202 & 0.248 &0.48 &7.99 &11.88 &21.81 &21.46 & 2.9&7.43& 02.6 \\ 
SDSS J005030.65+150134.0 & 12.62773 & 15.026127 & 6203-56266-898 & 0.409 &0.55 &14.13 &16.47 &20.65 &19.89 & 8.1&3.42& 13.4 \\ 
SDSS J014819.57+032424.9 & 27.08158 & 3.4069369 & 4272-55509-81 & 0.634 &0.94 &8.33 &10.12 &20.27 &20.10 & 8.7&8.94& 18.4 \\ 
SDSS J091318.66+604529.8 & 138.32775 & 60.758278 & 5712-56602-581 & 0.468 &0.84 &8.36 &12.05 &21.20 &20.01 & 5.2&10.61 & 9.7 \footnote{SDSS J0148+0324 does not meet the criteria for QSO-ELG selection, but exhibits doubled emission lines, as visible in figure \ref{QSOGal_candidates_plot}, probably due to two background galaxies. The flux reported above corresponds to only one emission line.}   \\ 
SDSS J092012.67-022230.4 & 140.05282 & -2.375115 & 3766-55213-95 & 0.657 &0.77 &14.77 &26.12 &21.68 &20.02 &19.6&16.53& 15.7 \\ 
SDSS J114059.35+000727.3 & 175.24732 & 0.12425373 & 3841-56016-740 & 0.430 &0.51 &8.32 &11.84 &20.55 &20.11 &14.2&9.01& 02.7 \\ 
SDSS J122905.05+215659.8 & 187.27105 & 21.949952 & 5982-56074-815 & 0.509 &0.81 &8.46 &11.34 &22.25 &20.29 & 6.6&9.41& 09.0 \\ 
SDSS J151020.36+185632.2 & 227.58485 & 18.942295 & 3951-55681-548 & 0.646 &0.73 &9.00 &11.69 &20.40 &19.76 &26.0&12.98& 23.7 \\ 
SDSS J233443.68+302159.9 & 353.68202 & 30.366664 & 6501-56563-228 & 0.079 &0.54 &10.33 &18.17 &19.33 &17.69 & 0.6&22.37& 17.6 \\ 
\hline
\end{tabular}
\tablefoot{Description of the fields: 1) SDSS target name in terms of truncated J2000 RA and DEC in the format HH.MM.SS.ss+DD .MM.SS.s 2) Right Ascension (RA) in degrees 3) DEClination (DEC) in degrees 4) Plate-MJD-Fiber of the spectrum 5) QSO redshift from the BOSS pipeline 6) Background galaxy redshift inferred from the emission lines detection 7) SN of the strongest emission line 8) Quadrature sum SN of the emission lines 9) g-band magnitude 10) i-band magnitude 11) Mass ($10^{12}M_\odot$)enclosed within the maximum detectable Einstein radius $\theta_E=3.6''$, assuming a Singular Isothermal Sphere (SIS) model and flat $\Lambda$CDM cosmology \citep{Planck2016}, giving an approximate upper bound on the lens mass. 12) Apparent flux (continuum subtracted) of the [OII]$\lambda\, 3727$ \AA\, emission line in units of $10^{-17}$ erg cm$^{-2}$ s$^{-1}$ 13) Apparent flux (continuum subtracted) of [OIII]$\lambda\, 5007$ \AA\, emission line in units of $10^{-17}$ erg cm$^{-2}$ s$^{-1}$}
\end{table}

\begin{table}
\caption{Selected properties of the QSO-LAE lens candidate systems }
\label{QSOLAE_candidates_BOSS}
\centering
\begin{tabular}{lrrcrrrrrrrrrr}
Target & RA & DEC & Plate-MJD-Fiber & $z_{l}$ &$z_s$ & $SN_{Ly\alpha}$  &$a_\lambda$  &$g$  & $i$ &$M_{SIS}$ &$f_{LyA}$ \\
\hline\hline
SDSS J013934.96+010629.9 & 24.895706 & 1.1083212 & 4231-55444-622 & 0.96  &4.01 & 9.82 & 2.24&22.41 &21.33 & 4.9 &8.80 \\
SDSS J101625.37+503427.0 & 154.10572 & 50.574194 & 6668-56605-342 & 1.23 &4.48 & 11.24 & 1.93&20.23 &19.67 & 5.8 &11.13 \\
SDSS J235422.48+195141.3 & 358.59368 & 19.861487 & 6110-56279-406 & 0.75 &2.17 & 10.66 & 0.21&18.64 &18.40 & 4.9 &72.31 \\
\hline
\end{tabular}
\tablefoot{Description of the fields: 1) SDSS target name in terms of truncated J2000 RA and DEC in the format HH.MM.SS.ss+DD.MM.SS.s 2) Right Ascension (RA) in degrees 3) DEClination (DEC) in degrees 4) Plate-MJD-Fiber of the spectrum for the BOSS target 5) QSO redshift from the BOSS pipeline 6) Background LAE redshift inferred from the peak detection 7) Signal-to-Noise ratio of the detected line 8) Wavelength ratio $a_\lambda$ \citep{Rhoads2003}, further described in section \ref{Lya_section} 9) g-band magnitude 10) i-band magnitude 11) Mass ($10^{12}M_\odot$)enclosed within the maximum detectable Einstein radius $\theta_E=3.6''$, assuming a Singular Isothermal Sphere (SIS) model and flat $\Lambda$CDM cosmology \citep{Planck2016}, giving an approximate upper bound on the lens mass. 12) Apparent flux (continuum subtracted) of the Lyman-$\alpha$ emission in units of $10^{-17}$ erg cm$^{-2}$ s$^{-1}$}
\end{table}
\end{landscape}

\section{Discussion}
\label{section_discussion}

\subsection{Selection function of our algorithm}
\label{sec:completeness}
\begin{figure*}
\centering 
\includegraphics[width =\textwidth]{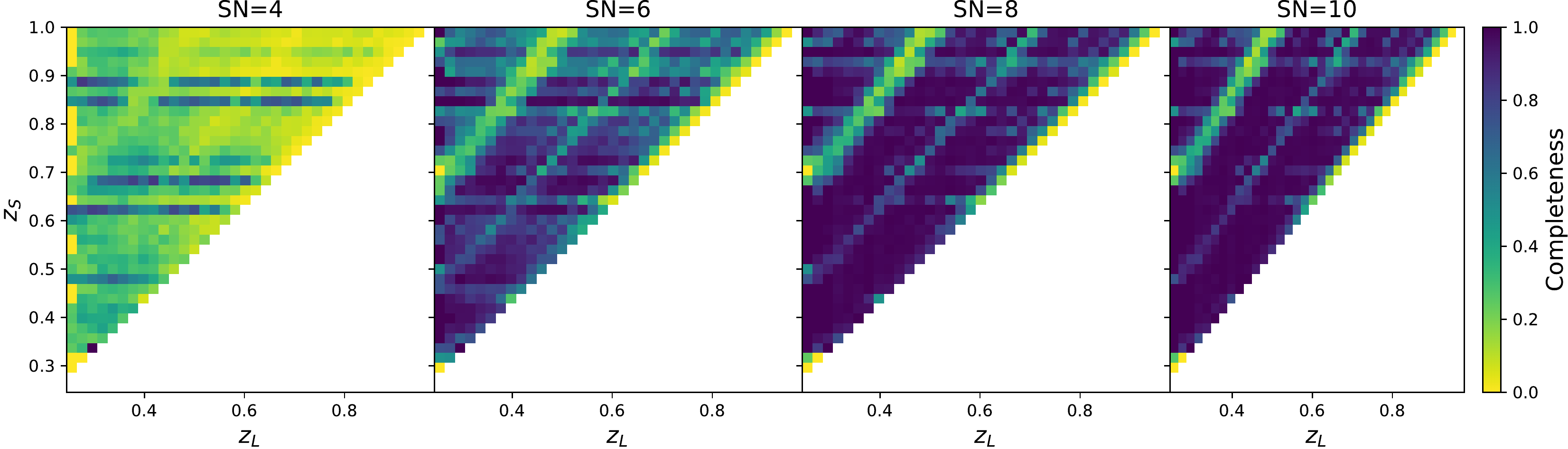}
\caption{Completeness function for the QSO-ELG lens system search with a detection threshold set at SN = 4,6,8,10 (left to right). The completeness rises quickly to $1$ above the set threshold and is only slightly impeded by the masking of broad and narrow QSO emission lines, as visible in linear trends of low completeness in the $(z_S,z_L)$ plot.  \label{fig:completeness_ELG}}
\end{figure*}

\begin{figure*}
\centering 
\includegraphics[width =\textwidth]{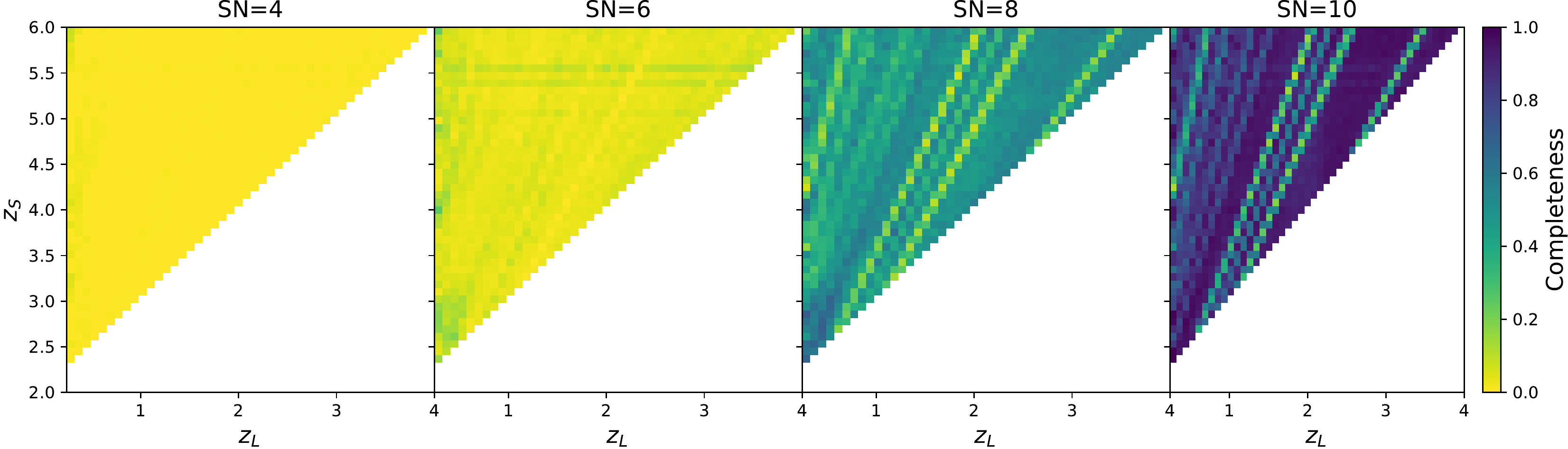}
\caption{ Completeness function for the QSO-SLE lens systems search with a detection threshold set at SN = 4,6,8,10 (left to right). The completeness is heavily impacted by the masking of broad and narrow QSO emission lines, as visible in linear trends of low completeness in the $(z_S,z_L)$ plot. \label{fig:completeness_SLE}}
\end{figure*}

We assess here the selection function of our algorithm, which is of crucial importance to any cosmological applications relying on the number density of quasars acting as lenses. We already noted above that the search is limited by the masking of the broad emission lines that are usually badly fitted by the PCA in the SDSS pipeline. 
The masked fraction  of the observed wavelength evolves from $\sim20$\% to $\sim50\%$ with redshift as the broader QSO UV lines enter the SDSS spectrum. Additionally, additional emission lines are virtually impossible to detect in the Lyman-$\alpha$ forest of the QSO, reducing even more the wavelength search range for QSO at $z\gtrsim2$. This severly limits the detection of higher-redshift lensed systems. As we show below (see Fig. \ref{fig:completeness_ELG} and \ref{fig:completeness_SLE}), the masking impacts more the detections of background LAEs than ELGs. The first reason is of course that ELGs are searched for at $z\lesssim 1$, where the masking is only about $\lesssim 20$\% of the observed wavelength range. The second is that ELGs can rely on five different lines distributed throughout the spectra, and so the masking of one does not prevent a successful detection.

We compute the completeness of our search by inserting mock ELG emission lines (only [OII], [OIII], H$\beta$ and H$\alpha$ at the same SN for simplicity), and mock SLEs in random SDSS spectra. We model the emission lines by Gaussian profiles with a variance of 1.2 pixels, rescaled to match a chosen SN. We divide the $(z_l,z_s)$ space in a fine grid for which we insert $1000$ emission features in randomly selected foreground QSOs spectra from DR12Q. The process is repeated for all SN in the range $[4,12]$. We present in Figures \ref{fig:completeness_ELG} and \ref{fig:completeness_SLE} the completeness of our algorithmic search for ELGs and SLEs, respectively. We first note the overall good performance of the search for ELGs, and its good completeness at $SN=8$ which was the threshold chosen for feature detection for this study. As said above, the ELG search is only lightly impeded by the broad line masking because it can use up to $5$ different lines to detect a background galaxy. The masking of QSO emission lines creates linear traces of low-completeness in the $(z_S,z_L)$ space. However, this affects more systems with $z_L>0.7$. Indeed, at $z_L>0.7$ [OIII] moves out of the allowed wavelength range ($<8500$ \AA) and thus candidates are retrieved only if both [OII] and H$\beta$ are detected. If one of them falls close to one of the broad QSO emission lines, the completeness drops close to zero. At $z<0.7$, this effect is mitigated since more than three lines can be detected and the chance of haveing them all fall in masked regions of the spectra is virtually null.

The SLE completeness is about a factor $30$ to $50$\% lower compared to ELG at the same SN, which cannot only be attributed to a larger sensitivity to the masking. Indeed, all ELG features are masked only for specific combinations of $(z_S, z_L)$, whereas single emitter lines are never detected if they fall in some of the masked regions, which can make up a large fraction of the spectra. The remaining drop in efficiency is easily explained by the scatter in the estimated SN and the real SN. When multiple lines are used, the probability that all of them are underestimated due to continuum residuals is low, whereas an underestimated SLE close to the SN threshold is always rejected. 

The purity of the algorithmic search requires a comprehensive modelling of QSO spectra and background sources and is thus beyond the scope of this paper. The human confirmation process of course tends to increase purity and diminish completeness,  but we are confident that this process is negligible for the QSO-ELG lenses sample. Only a small number of systems ($\sim 250$) had to be checked by eye for ELG features, and the nature of false positives was always obvious (e.g. clear continuum residuals or clipped spectrum). The selection of QSO-LAE lenses is more difficult because measuring the asymmetry of the profile is difficult at the resolution of the BOSS spectrograph where low-SN features are not well-resolved. We have attempted to classify SLEs based on different asymmetry indicators such as skewness and the  \citet{Rhoads2003} wavelength ratio, but they fail to separate clearly candidates selected by eye from the parent SLE sample. We note that no QSO-LAE lens system has ever been observed to date, thus a first detection would motivate a fully automated selection from spectrosocopic surveys of such lensing systems.

We have hence characterised our selection function, showing that the search for background ELG is complete at the $>90$\% for the chosen SN threshold, and that the QSO-LAE search is mostly impeded by the masking of broad emission lines. In the latter case, a more careful modelling of QSO continua could boost the number of detections by a factor $\sim 2$, which could be very significant given the prospect of future large spectroscopic samples of QSOs.

\subsection{Number of QSO-ELG systems}

We now discuss the number of candidates obtained through our search across all SDSS-III QSOs, and compare it to the only previous search for such lenses.
We first draw some parallels and differences between \citet{Courbin2012} first $3$ confirmed QSO-galaxy lenses and our $9$ candidates. The two candidates samples do not overlap because we did not apply our algorithm to SDSS-I/-II.Indeed, our automated method is built around the PCA template for quasars which was introduced in SDSS-III. It must be noted that all the QSO-ELG lenses shortlisted in Table \ref{QSOGal_best} have a lens redshift $z<0.7$, as did the SDSS-II QSOs in \citet{Courbin2010}. Our candidate selection is based on the detection of multiple emission lines among [OII], H$\beta$, [OIII] and H$\alpha$. Even though H$\alpha$ is not essential for a detection, [OII], [OIII] and H$\beta$ should be clearly visible. This sets a detection limit for background galaxies lensed by QSOs, i.e. $z\lesssim 0.8$, as [OIII] and H$\beta$ at higher redshift are off the red end of the SDSS spectrograph. As the red-end of the BOSS spectrograph is often too noisy and affected by strong sky emission, only $z\lesssim0.7$ QSOs can in fact be used to detect QSO-ELGs systems. BOSS primarily targeted QSOs at redshift $z>2$. Hence we are limited by the number of QSOs in the low-redshift tail of the distribution of BOSS QSOs, or objects re-observed from SDSS-I/II. The number of QSOs with redshift $z<0.7$ is $31081$ in BOSS DR12 and thus we reach a fraction of selected QSO-ELG lens candidates of $9/31081 \approx 3 \times 10^{-4}$. 

\subsection{Future perspectives}
We have demonstrated the potential of our spectroscopic selection of QSO lensing background ELGs/LAEs where image-based techniques often fail due to the brightness of the foreground target. In the case of QSO-ELG lenses, the presence of at least $4$ emission lines at a total SN of $\gtrsim 11\sigma$ guarantees the presence of a higher-redshift background galaxy. With high spatial resolution follow-up, this sample may increase the number of such QSO lenses by a factor of $3$-$4$. 

QSO-LAE lens candidates are more tentative candidates. However, there is hope to confirm the first ever QSO-LAE strong lensing system. The remaining SLEs sample could contain more QSO-LAE candidates, but there is not enough evidence to draw firm conclusions only based on the spectra. We note however that selecting galaxies on Lyman$-\alpha$ only was succussful for the BELLS-GALLERY survey. Because Lyman-$\alpha$ can be detected in optical spectra at a wider range of redshifts than nebular emission lines, most QSO acting strong lenses could be in the future QSO-LAE systems. Improving the modelling of the QSO continuum will reduce the masked fraction of the observed spectra, potentially doubling the wavelength search space and thus the number of candidates. Confirming the first QSO-LAE system is thus the next step in the direction of statistically meaningful sample of QSO acting as strong gravitational lenses. 
This is of prime importance for the applications of large samples: constraining the $M_{BH}$-$\sigma_*$ relationship from the QSO broadened lines and the mass inferred from the lens modelling, as well as using the observed lensing probabilities to differentiate halo models of QSOs.

\begin{acknowledgements}
JPK, TD acknowledge support from the ERC advanced grant LIDA. FC acknowledges support from the  Swiss National  Science Foundation (SNSF). We thank an anonymous referee for comments on a first version of this paper which significantly improved it. We thank Mathilde Jauzac for useful feedback on an early draft of this manuscript. We are grateful to the authors of the following Python packages: Matplotlib \citep{Matplotlib}, Numpy \citep{Numpy} and Scipy \citep{Scipy}.
\\
Funding for SDSS-III has been provided by the Alfred P. Sloan Foundation, the Participating Institutions, the National Science Foundation, and the U.S. Department of Energy Office of Science. The SDSS-III web site is http://www.sdss3.org/.

SDSS-III is managed by the Astrophysical Research Consortium for the Participating Institutions of the SDSS-III Collaboration including the University of Arizona, the Brazilian Participation Group, Brookhaven National Laboratory, Carnegie Mellon University, University of Florida, the French Participation Group, the German Participation Group, Harvard University, the Instituto de Astrofisica de Canarias, the Michigan State/Notre Dame/JINA Participation Group, Johns Hopkins University, Lawrence Berkeley National Laboratory, Max Planck Institute for Astrophysics, Max Planck Institute for Extraterrestrial Physics, New Mexico State University, New York University, Ohio State University, Pennsylvania State University, University of Portsmouth, Princeton University, the Spanish Participation Group, University of Tokyo, University of Utah, Vanderbilt University, University of Virginia, University of Washington, and Yale University. 
\end{acknowledgements}

\bibliographystyle{aa}
\bibliography{QSOLens}

\begin{appendix}
\onecolumn

\section{QSO-SLE detections}
\begin{figure}
\centering
\includegraphics[width = 0.95\textwidth]{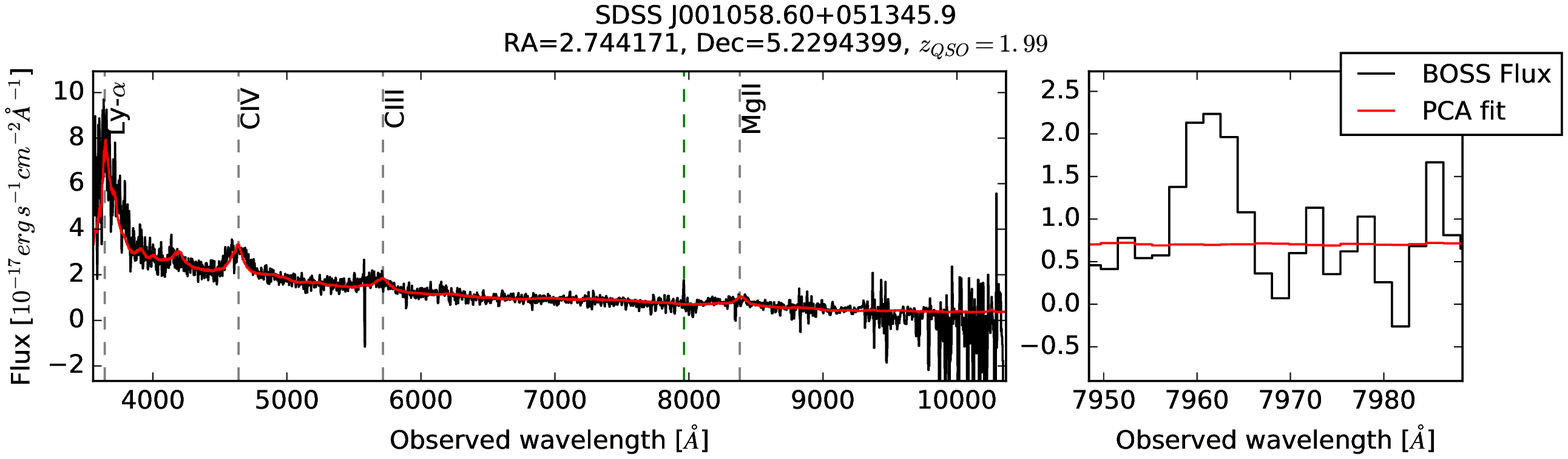} \\
\includegraphics[width = 0.95\textwidth]{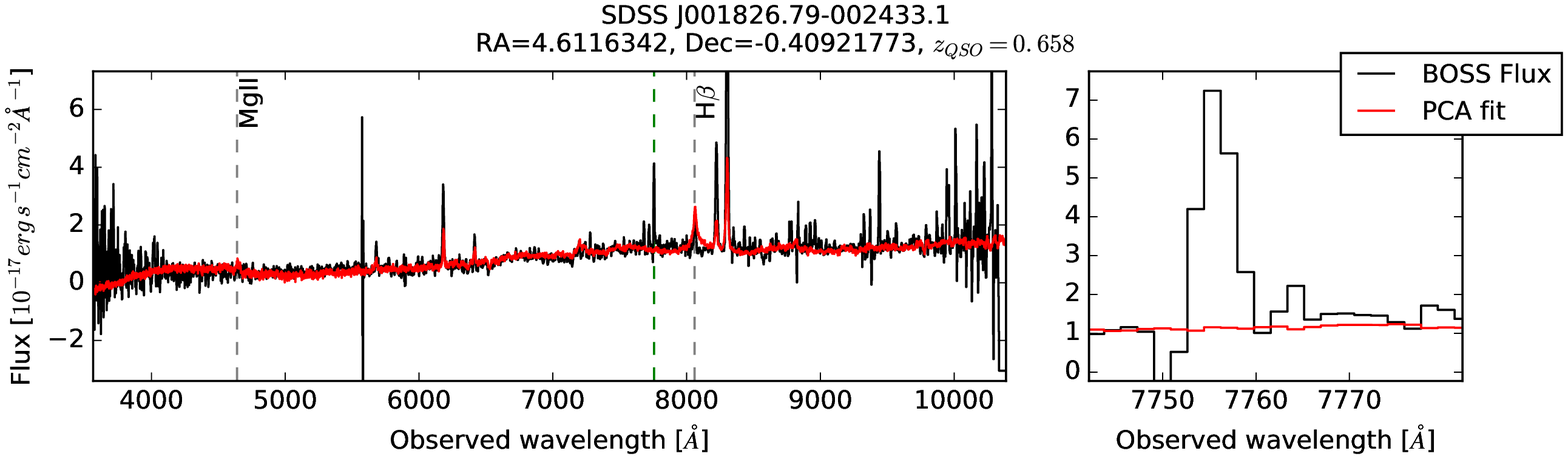} \\
\includegraphics[width = 0.95\textwidth]{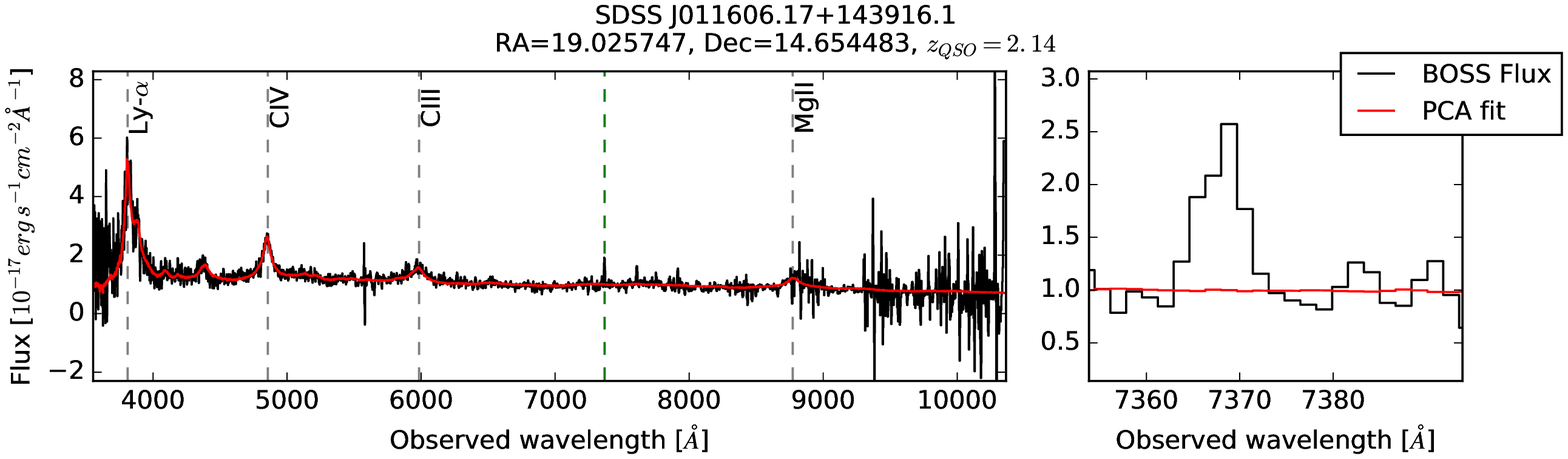} 
\caption{SDSS spectra of the 3 first SLEs of the sample. Even though a clear emission line is detected in each QSO spectrum. Emission lines are either not asymmetric enough to qualify as a potential Lyman-$\alpha$ or appear in noisy regions of the spectra.}
\label{fig:SinglelineEmitters}
\end{figure}

We present the three first SLEs from the sample of $49$ that were not deemed sufficiently asymetric to be considered QSO-LAE candidates. Figure \ref{fig:SinglelineEmitters} shows the QSO spectra and the emission line from the SLE. The distribution of the source and lens redshifts for the QSO-SLEs systems is given in Figure \ref{histo}, where the redshift of the SLE is computed assuming the emission line is indeed Lyman-$\alpha$ or alternatively O{~\small II} 3727 \AA.
\label{appendix:SLE}

\section{Masked QSO emission and sky emission lines }
We report the masked regions of the observed SDSS spectra due to QSO emission lines (Table \ref{mask_table_QSO}) and common sky lines \ref{mask_table_sky}. In each case, we give the central rest-frame wavelength and the rest-frame width of the masked region in Angstr\"{o}ms. Note that the observed width of the masked regions increases with redshift and diminishes the search space for supplementary emission lines, as discussed in Section \ref{section_discussion}.

\begin{table}
\caption{Masked QSO emission lines (see Section \ref{section_method})}
\label{mask_table_QSO} 
\centering
\begin{tabular}{lrr}
Denomination & Wavelength [\AA] & Mask width [\AA] \\
\hline \hline
Lyman-$\alpha$ & 1215.57 & 300 \\
N V 1240 & 1240.81 & 25 \\
SiIV + OIV 1400 & 1399.80 & 75 \\
C IV 1549 & 1549.48  & 100 \\
He II 1640 & 1640.42 & 50 \\
C III] 1908 & 1908.73 	& 50 \\
C II 2326 & 2326 & 25 \\
Ne IV 2423 & 2423.83 & 50 \\
Mg II 2799 & 2799.49  & 125 \\
Ne V 3347 & 3346.79 & 25 \\
Ne V 3427 & 3426.85 & 25 \\
\text{[OII]} 3727 & 3727 & 50 \\
\text{[Ne III]} 3868 & 3868.75 & 25 \\
H$\delta$ & 4101.73 & 25 \\
H$\gamma$ & 4340.46 & 75 \\
Weak Fe & 4490 & 50 \\
H$\beta$ & 4861.325 & 50 \\
\text{[O III]} 4959 & 4958.91 & 100 \\
\text{[O III]} 5007 & 5006.84 & 100 \\
Weak Fe & 5080 & 50\\
N I 5200 & 5200.53 & 50 \\
Weak Fe & 5317 & 50 \\
Weak Fe & 5691 &  50\\
\text{[Fe VII]} 5722 & 5722.30 & 25 \\
\text{[Fe VII]} 6087 & 6087.98 & 25 \\
Weak Fe & 6504 & 50 \\
H$\alpha$ & 6562.80 & 125 \\
\text{[S II]} 6716 & 6716.44 & 50 \\
\text{[S II]} 6730 & 6730.82 & 50 \\
\hline
\end{tabular}
\tablefoot{Masked emission lines and width of the mask. Due to the large widths and the binning of the spectra, the central wavelength is rounded up for the implementation. The most important lines were masked \textit{a priori}, while the fainter iron features were added \textit{a posteriori} following the over-densities of false positives in the QSO rest-frame}
\end{table}

\begin{table}
\caption{Masked sky emission lines  (see Section \ref{section_method})}
\label{mask_table_sky}
\centering
\begin{tabular}{lrr}
Denomination & Wavelength [\AA] & Mask width [\AA] \\
\hline \hline
Sky 5577 & 5577 & 20 \\
Sky 5896 & 5895 & 25 \\
Sky 6300 & 6300 & 30 \\
Sky 6363 & 6363 & 30 \\
\end{tabular}
\end{table}
\clearpage
\end{appendix}
\end{document}